\documentclass[12pt]{article}

\usepackage[utf8]{inputenc}
\usepackage{bbm}
\usepackage{bm}
\usepackage{ragged2e}
\usepackage{mathrsfs}
\usepackage{amsfonts}
\usepackage{float}
\usepackage{amsthm}
\usepackage{epsfig,rotating}
\usepackage{amssymb,amsmath}
\usepackage{latexsym}
\usepackage{graphicx}
\usepackage{caption,subcaption}
\usepackage{multirow}
\usepackage{booktabs,natbib}
\usepackage{comment}
\usepackage{graphicx}
\graphicspath{{Graph/}}
\usepackage{xcolor}
\usepackage{diagbox}
\usepackage{comment}
\usepackage{tabularx}
\usepackage{algorithmicx}
\usepackage[ruled,vlined]{algorithm2e}

\usepackage{geometry}
\usepackage{multibib}
\usepackage{endnotes}
\usepackage[titletoc]{appendix}
\bibpunct{(}{)}{;}{a}{,}{,}
\newcites{New}{References}

\newtheorem{definition}{\noindent D{\footnotesize EFINITION}}
\newtheorem{assumption}{\noindent A{\footnotesize SSUMPTION}}
\newtheorem{theorem}{\noindent T{\footnotesize HEOREM}}
\newtheorem{proposition}{\noindent P{\footnotesize ROPOSITION}}

\newtheorem{lemma}{\noindent L{\footnotesize EMMA}}
\newtheorem{coro}{ \noindent C{\footnotesize OROLLARY}}
\newtheorem{remark}{\noindent R{\footnotesize EMARK}}

\newcommand{\RN}[1]{%
  \textup{\uppercase\expandafter{\romannumeral#1}}%
}

\newcommand{\Rbb}{\mathbb{R}}

\newcommand{\Ffr}{\mathfrak{F}}

\newcommand{\Bcal}{\mathcal{B}}
\newcommand{\Lcal}{\mathcal{L}}
\newcommand{\Hcal}{\mathcal{H}}

\newcommand{\Fcal}{\mathcal{F}}
\newcommand{\Pcal}{\mathcal{P}}

\newcommand{\Scal}{\mathcal{S}}

\def\nano{\scriptscriptstyle}
\newcommand\hi[1]{^{\nano #1}}
\def\real{\mathbb R}
\newcommand\ca[1]{{\cal{#1}}}
\newcommand\lo[1]{_{\nano #1}}

\def\cov{\mathrm{cov}}
\def\nano{\scriptscriptstyle}
\def\inv{\hi{\nano -1}}

\def\nano{\scriptscriptstyle}

\def\ka{\kappa}

\def\tsum{\textstyle{\sum}}

\def\diag{\mathrm{diag}}

\def\hii#1{\hi{(#1)}}

\def\spn{\mathrm{span}}

\def\diag{\mbox{diag}}

\newcommand{\trans}{^{\mbox{\tiny {\sf T}}}}

\newcommand{\indep}{\rotatebox[origin=c]{90}{$\models$}}
\newcommand{\tr}{\mathrm{tr}}
\newcommand{\Sym}{\mathrm{Sym}}

\newcommand{\W}{\mathcal{W}_2(I)}

\newcommand{\var}{\mathrm{var}}

\makeatletter
\def\old@comma{,}
 \catcode`\,=13
 \def,{%
   \ifmmode%
     \old@comma\discretionary{}{}{}%
   \else%
     \old@comma%
   \fi%
 }
 \makeatother

\title{\Large Dimension Reduction for Fr\'echet Regression}
\author{
\normalsize Qi Zhang, Lingzhou Xue, and Bing Li\\ \normalsize Department of Statistics, Pennsylvania State University
}
\date{}

\begin{document}

\maketitle

\begin{abstract}
With the rapid development of data collection techniques, complex data objects that are not in the Euclidean space are frequently encountered in new statistical applications. Fr\'echet regression model (Peterson \& M\"uller 2019) provides a promising framework for regression analysis with metric space-valued responses. In this paper, we introduce a flexible sufficient dimension reduction (SDR) method for Fr\'echet regression to achieve two purposes: to mitigate the curse of dimensionality caused by high-dimensional predictors and to provide a { visual inspection tool for Fr\'echet regression}. Our approach is flexible enough to turn any existing SDR method for Euclidean $(X,Y)$ into one for Euclidean $X$ and metric space-valued $Y$. The basic idea is to first map the metric space-valued random object $Y$ to a real-valued random variable $f(Y)$ using a class of functions and then perform classical SDR to the transformed data. If the class of functions is sufficiently rich, then we are guaranteed to uncover the Fr\'echet SDR space. We showed that such a class, which we call an ensemble, can be generated by a universal kernel (cc-universal kernel). We established the consistency and asymptotic convergence rate of the proposed methods. The finite-sample performance of the proposed methods is illustrated through simulation studies for several commonly encountered metric spaces that include Wasserstein space, the space of symmetric positive definite matrices, and the sphere. We illustrated the data visualization aspect of our method by the human mortality distribution data from the United Nations Databases.
\end{abstract}



\section{Introduction}
With the rapid development of data collection techniques, complex data objects that are not in the Euclidean space are frequently encountered in new statistical applications, such as the graph Laplacians of networks, the covariance or correlation matrices for the brain functional connectivity in neuroscience \cite{ferreira2013resting}, and probability distributions in CT hematoma density data \citep{petersen2019frechet}. These data objects, also known as random objects, do not obey the operation rules of a vector space with an inner product or a norm but instead reside in a general metric space. In a prescient paper, \cite{frechet1948elements} proposed the Fr\'echet mean as a natural generalization of the expectation of a random vector. By extending the Fr\'echet mean to the conditional Fr\'echet mean, \cite{petersen2019frechet} introduced the Fr\'echet regression model with random objects as the response and Euclidean vectors as predictors, which provides a promising framework for regression analysis with metric space-valued responses. \cite{dubey2019frechet} showed the consistency of the sample Fr\'echet mean using the results of \cite{petersen2019frechet}, derived a central limit theorem for the sample Fr\'echet variance that quantifies the variation around the Fr\'echet mean, and further developed the Fr\'echet analysis of variance for random objects. \cite{dubey2020frechet} designed a method for change-point detection and inference in a sequence of metric-space-valued data objects. 

The Fr\'echet regression of \cite{petersen2019frechet} employs the global least squares and the local linear or polynomial regression to fit the conditional Fr\'echet mean. It is well known that the global least squares are based on a restrictive assumption of the regression relation. Although the local regression is more flexible, it is effective only when the dimension of the predictor is relatively low. As this dimension gets higher, its accuracy drops significantly--a phenomenon known as the curse of dimensionality. To address this issue, it is essential to reduce the dimension of the predictor without losing the information about the response. For classical regression, this task is performed by sufficient dimension reduction (SDR; see \citealt{li1991sliced,cook1991sliced,cook1996graphics} and \citealt{li2018sufficient} among others). SDR works by projecting the high-dimensional predictor onto a low-dimensional subspace that preserves the information about the response through the use of sufficiency.

Besides assisting regression in overcoming the curse of dimensionality, another important function of SDR for classical regression is to provide a data visualization tool to gain insights into how the regression surface looks in high-dimensional space before even fitting a model. { By inspecting the sufficient plots of the response objects against the sufficient predictors, we can gain insights into the general trends of the response as the most informative part of the predictor varies, whether there are outlying observations, and whether there are subjects with high leverage that have undue influence on the regression estimates-the usual information a statistician looks for in the exploratory and model checking stages of the regression analysis.} This function is also needed in Fr\'echet regression. In fact, it can be argued that data visualization is even more important for the regression of random objects, as the regression relation may be even more difficult to discern among the complex details of the objects.

To fulfill these demands, in this paper, we systematically develop the theories and methodologies of sufficient dimension reduction for Fr\'echet regression. To set the stage, we first give an outline of SDR for classical regression. Let $X$ be a $p$-dimensional random vector in $\mathbb{R}^p$ and
$Y$ a random variable in $\mathbb{R}$. The classical SDR aims to find a dimension reduction subspace $\Scal$ of $\Rbb^p$ such that $Y$ and $X$ are independent conditioning on $P_{\Scal}X$, that is,
$
Y\indep X|P_{\Scal}X,
$
where $P_{\Scal}$ is the projection on to $\Scal$ with respect to the usual inner product in $\Rbb^p$. In this way, $P_{\Scal}X$ can be used as the synthetic predictor without loss of regression information about the response $Y$. Under mild conditions, the intersection of all such dimension reduction  subspaces is also a dimension reduction subspace, and the intersection is called the central subspace denoted by $\Scal_{Y|X}$ \citep{cook1996graphics,yin2008successive}. For the situation where the primary interest is in estimating the regression function, \cite{cook2002dimension}
introduced a weaker form of SDR, the mean dimension reduction subspace. A subspace $\Scal$ of $\Rbb^p$ is a mean SDR subspace if satisfies $E(Y|X) = E(Y|P_{\Scal}X)$, and the intersection of all such spaces if it is still a mean SDR subspace, is the central mean subspace, denoted by  $\Scal_{E(Y|X)}$. The central mean subspace  $\Scal_{E(Y|X)}$ is always contained in central subspace $\Scal_{Y|X}$ when they exist.  Many estimating methods for the central subspace and the central mean subspace have been developed over the past decades. For example, for the central subspace, we have the sliced inverse regression (SIR; \citealt{li1991sliced}), the sliced average variance estimate (SAVE; \citealt{cook1991sliced}), the contour regression (CR; \citealt{li2005contour}), and the directional regression (DR; \citealt{li2007directional}). For the central mean subspace, we have the ordinary least squares (OLS; \citealt{li1989regression}), the principal Hessian directions (PHD; \citealt{li1992principal}), the iterative Hessian transformation (IHT \citealt{cook2002dimension}), the outer product of gradients (OPG) and the minimum average variance estimator (MAVE) of \cite{xia2002adaptive}.

SDR has been extended to accommodate some complex data structures in the past, for example, to functional data (\citealt{ferre2003functional}; \citealt{hsing2009rkhs}; 
\citealt{li2017nonlinear}), to tensorial data (\citealt{li2010dimension}; \citealt{ding2015tensor}), and to panel data \citep{fan2017sufficient,yu2020nonparametric,luo2021inverse}. Most recently, \cite{ying2020fr} extended SIR to the case where the response takes values in a metric space. Taking a substantial step forward, in this paper, we introduce a comprehensive and flexible method that can adapt any existing SDR estimators to metric space-valued responses.

The basic idea of our method stems from the ensemble SDR for Euclidean $X$ and $Y$ of \cite{yin2011sufficient}, which recovers the central subspace $\Scal_{Y|X}$ by repeatedly estimating the central mean subspace $\Scal_{E[f(Y)|X]}$ for a family $\Ffr$ of functions $f$ that is rich enough to determine the conditional distribution of $Y|X$. Such a family $\Ffr$ is called an ensemble and satisfies
$
\Scal_{Y|X} = \cup \{ \Scal_{E[f(Y)|X]}: f \in \Ffr \}.
$
Using this relation, we can turn any method for estimating the central mean space into one that estimates the central subspace.

While borrowing the idea of the ensemble, our goal is different from \cite{yin2011sufficient}: we are not interested in turning an estimator for the central mean subspace into one for the central subspace. Instead, we are interested in turning any existing SDR method
for Euclidean $(X,Y)$ into one for Euclidean $X$ and metric space-valued $Y$. Let $X$ be a random vector in $\Rbb^p$ and $Y$ a random object that takes values in a metric space $(\Omega_Y, d)$. Still use the symbol $S_{Y|X}$ to represent the intersection of all subspaces of $\Rbb^p$ satisfying $Y\indep X|P_{\Scal} X$. We call $\Scal\lo {Y|X}$ the central subspace for Fr\'echet SDR, or simply the Fr\'echet central subspace. Let $\Ffr$ be a family of functions $f: \Omega_Y \to \Rbb$ that are measurable with respect to the Borel $\sigma$-field on the metric space. We use two types of ensembles to connect classical SDR with Fr\'echet SDR:
\begin{itemize}
\item Central Mean Space ensemble (CMS-ensemble) is a family $\Ffr$ that is rich enough so that
$
\Scal_{Y|X} = \cup \{ \Scal_{E[f(Y)|X]}: f \in \Ffr \}.
$
Note that we know how to estimate the spaces $\Scal_{E(f(Y)|X)}$ using the existing SDR methods since $f(Y)$ is a number. We use this ensemble to turn an SDR method that targets the central mean subspace into one that targets the Fr\'echet central subspace. We will focus on two forward regression methods: OPG and MAVE, and three moment estimators of the CMS.
\item Central Space ensemble (CS-ensemble) is a family $\Ffr$ that is rich enough so that
$
\Scal_{Y|X} = \cup \{ \Scal_{f(Y)|X}: f \in \Ffr \}.
$
We use this ensemble to turn an SDR method that targets the central subspace for real-valued response into one that targets the Fr\'echet central subspace. We will focus on three inverse regression methods: SIR, SAVE, and DR.
\end{itemize}
A key step in implementing both of the above schemes is to construct an ensemble $\Ffr$ in each case. For this purpose, we assume that the metric space $(\Omega_Y, d)$ is continuously embeddable into a Hilbert space. Under this assumption, one can construct a universal reproducing kernel, which leads to an $\Ffr$ that satisfies the required characterizing property.

\begin{figure}[ht!]
	\centering
	\begin{subfigure}[b]{0.35\textwidth}
         \centering
         \includegraphics[width=\textwidth]{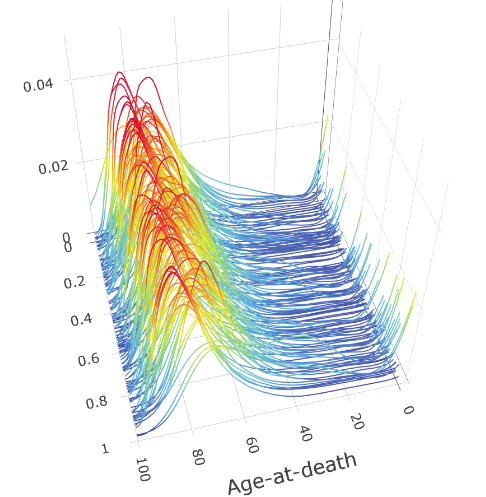}
         \caption{}
     \end{subfigure}
     \begin{subfigure}[b]{0.35\textwidth}
         \centering
         \includegraphics[width=\textwidth]{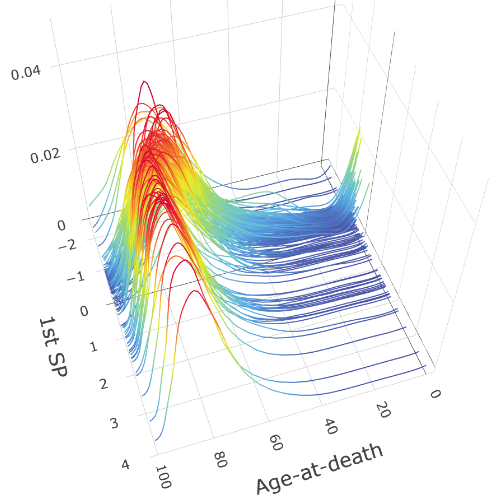}
         \caption{}
     \end{subfigure}
	\caption{Data visualization in Fr\'echet regression for mortality distributions of 162 countries. Panel (a) plots mortality densities that are placed in random order, and Panel (b) plots mortality densities versus the first sufficient predictor estimated by our ensemble method.}\label{fig:1}
\end{figure}

As with classical SDR, the Fr\'echet SDR can also be used to assist data visualization. To illustrate this aspect, we consider an application involving factors that influence the mortality distributions of 162 countries  (see Section \ref{sec:realdata} for details). For each country, the response is a histogram with the numbers of deaths for each five-year period from age 0 to age 100, which is smoothed to produce a density estimate, as shown in panel (a) of Figure \ref{fig:1}. We considered nine predictors characterizing each country's demography, economy, labor market, health care, and environment. Using our ensemble method we obtained a set of sufficient predictors. In panel (b) of Figure \ref{fig:1}, we show the mortality densities plotted against the first sufficient predictor. A clear pattern is shown in the plot: for countries with low values of the first sufficient predictor, the modes of the mortality distributions are at lower ages, and there are upticks at age 0, indicating high infant mortality rates; for countries with high values of the first sufficient predictor, the modes of the mortality distributions are significantly higher, and there are no upticks at age 0, indicating very low infant mortality rates. The information provided by the plot is clearly useful, and many further insights can be gained about what affects the mortality distribution by taking a careful look at the loadings of the first sufficient predictor, as will be detailed in Section \ref{sec:realdata}.

The rest of this paper is organized as follows.  Section \ref{sec:fcs}  defines the Fr\'echet SDR problem and provides sufficient conditions for a family $\Ffr$ to characterize the central subspace. Section 3 then constructs ensemble $\Ffr$ for the Wasserstein space of univariate distributions, the space of covariance matrix, and a special Riemannian manifold, the sphere. Section \ref{sec:fsdr} proposes the CMS-ensembles by extending five SDR methods that target the central mean space for real-valued response: OLS, PHD, IHT, OPG and MAVE, and CS-ensembles by extending three SDR methods that target the central space for real-valued response: SIR, SAVE, and DR. Section \ref{sec:asymptotics} establishes the convergence rate of the proposed methods. Section \ref{sec:simu} uses simulation studies to examine the numerical performances of different ensemble estimators in different settings, including distributional responses and covariance matrix responses. In Section \ref{sec:realdata}, we analyze the mortality distribution data to demonstrate the usefulness of our methods. Section \ref{sec:discuss} includes a few concluding remarks and discussion. All the proofs and additional simulation studies and real application are presented in the Supplementary Material.

\section{Characterization of the Fr\'echet Central Subspace}\label{sec:fcs}
Let $(\Omega,\Fcal, P)$ be a probability space. Let $(\Omega_Y, d)$ be a metric space with metric $d$ and $\mathcal{B}_Y$ the Borel $\sigma$-field generated by the open sets in $\Omega_Y$. Let $\Omega_X$ be a subset of $\Rbb^p$ and $\mathcal{B}_X$ the Borel $\sigma$-field generated by the open sets in $\Omega_X$. Let $(X,Y)$ be a random element mapping from $\Omega$ to $\Omega_X\times \Omega_Y$ measurable with respect to the product $\sigma$-field $\Bcal_X\times\Bcal_Y$.
We denote the marginal distributions of $X$ and $Y$ by $P_X$ and $P_Y$, respectively, and the conditional distributions of $Y|X$ and $X|Y$ by $P_{Y |X}$ and $P_{X|Y}$, respectively. We formulate the Fr\'echet SDR problem as finding a subspace $\Scal$ of $\Rbb^p$ such that $Y$ and $X$ are independent conditioning on $P_{\Scal}X$:
\begin{equation}\label{eq:sdrdef}
    Y\indep X|P_{\Scal}X,
\end{equation}
where $P_{\Scal}$ is the projection on to $\Scal$ with respect to the inner product in $\Rbb^p$. As in the classical SDR, the intersection of all such subspaces $\Scal$ still satisfies \eqref{eq:sdrdef} under mild conditions \citep{cook2002dimension}. { Indeed, it does not require any structure of the space $\Omega\lo Y$. A sufficient condition shown in \cite{yin2008successive} is that $X$ is supported by a matching set. For example, if the support of $X$ is convex, then this sufficient condition is satisfied. }We call this subspace the Fr\'echet central subspace and denote it by $\Scal_{Y|X}$. Similar to \cite{cook1996graphics}, it can be shown that if the support of X is open and convex, the Fr\'echet central subspace $\Scal_{Y|X}$ satisfies \eqref{eq:sdrdef}.

\subsection{Two types of ensembles and their sufficient conditions}
Let $\Ffr$ be a family of measurable functions $f :\Omega_Y \to \Rbb$, and for an $f\in\Ffr$, let $\Scal_{E[f(Y)|X]}$ be the central mean subspace of $f(Y)$ versus $X$. As mentioned in Section 1, we use two types of ensembles to recover the Fr\'echet central subspace. The first type is any  $\Ffr$ that satisfies
\begin{align}\label{eq:cmsens}
    {\spn}\{\Scal_{E(f(Y)|X)}: f\in\Ffr\}=\Scal_{Y|X}.
\end{align}
This is the same ensemble as that in \cite{yin2011sufficient}, except that, here, the right-hand side is the Fr\'echet central subspace. Relation \eqref{eq:cmsens} allows us to recover the Fr\'echet central subspace $\Scal_{Y|X}$ by a collection of classical central mean subspaces. We call a class $\Ffr$ that satisfies \eqref{eq:cmsens} a CMS-ensemble. The second type of ensembles is any family $\Ffr$ that satisfies
\begin{align}\label{eq:csens}
\spn \{ \Scal_{f(Y)|X}: \, f \in \Ffr \} = \Scal_{Y|X},
\end{align}
which we call a CS-ensemble. Proposition \ref{prop:cms-cs} shows that a CMS ensemble is  a CS-ensemble.
\begin{proposition} \label{prop:cms-cs}
If $\Ffr$ is a CMS-ensemble, then it is a CS-ensemble.
\end{proposition}

We next develop a sufficient condition for an $\Ffr$ to be a CMS-ensemble and hence also a CS-ensemble. Let $\mathfrak{B}=\{I_B: B\ \text{is Borel set in}\ \Omega_Y\}$ be the family of measurable indicator functions on $\Omega_Y$, and let $\spn(\Ffr)=\left\{\sum_{i=1}^k\alpha_if_i: k\in \mathbb{N}, \alpha_1,\dots,\alpha_k\in \Rbb, f_1,\dots,f_k\in \Ffr\right\}$ be the linear span of $\Ffr$, where $\mathbb N = \{1, 2, \dots\}$.
\cite{yin2011sufficient} showed that if $\Ffr$ is a subset of $L_2(P_Y)$ that is dense in $\mathfrak{B}$, then \eqref{eq:cmsens} holds for the classical $\Scal_{Y|X}$. Here, we generalize that result to our setting by requiring only $\spn(\Ffr)$  to be dense in $\mathfrak{B}$.
\begin{lemma}\label{lemma:charac}
    If $\Ffr$ is a subset of $L_2(P_Y)$ and $\spn\{\Ffr\}$ is dense in $\mathfrak{B}$ with respect to the $L_2(P_Y)$-metric, then $\Ffr$ is a CMS-ensemble and hence also a CS-ensemble.
\end{lemma}
\subsection{Construction of the CMS-ensemble}
To construct a CMS-ensemble, we resort to the notion of the \textit{universal kernel}. Let $C(\Omega_Y)$ be the family of continuous real-valued functions on $\Omega_Y$.  { When $\Omega\lo Y$ is compact, \citet{steinwart2001influence} defined a continuous kernel $\kappa$ as universal (we refer to it as \emph{c-universal}) if its associated RKHS $\ca H\lo Y$ is dense in $C(\Omega\lo Y)$ under the uniform norm. To relax the compactness assumption, \citet{micchelli2006universal} proposed the following notion of universality, which is referred to \emph{cc-univsersal} in \citet{sriperumbudur2011universality}.  For any compact set $K\subseteq\Omega\lo Y$, let $\ca H \lo Y (K)$ be the RKHS generated by $\{\ka (\cdot, y): y \in K \}$. We should note that a member $f$ of $\ca H \lo Y (K)$ is supported on $\Omega \lo Y$, rather than $K$. Let $f|K$ denote the restriction of $f$ on $K$, and $C(K)$ the class of all continuous functions with respect to the topology in $(\Omega \lo Y, d)$ restricted on $K$.
\begin{definition}\label{definition:universal}\citep{micchelli2006universal} We say that $\ka$ is cc-universal if, for any compact set $K \subseteq \Omega \lo Y$, any member $f$ of $C(K)$, and any $\epsilon > 0$, there is an $h \in \ca H \lo Y (K)$ such that
$
\|f - (h|{K})\|\lo\infty = \sup \lo {y \in K} | f (y) - h (y) | < \epsilon.
$
\end{definition}
When $\Omega\lo Y$ is compact, \citet{sriperumbudur2011universality} showed that two notions of universality are equivalent. In the following, we look into the conditions under which a metric space has a cc-universal kernel and how to construct such a kernel when it does.
}

\def\nat{\mathbb{N}}

\citet{micchelli2006universal} showed that when $\Omega\lo Y = \Rbb^d$, many standard kernels, including Laplacian kernels and Gaussian RBF kernels, are cc-universal. Unfortunately, when $\Omega_Y$ is a general metric space, direct extension of these types of kernels, for example, $k(y,y^\prime)=\exp(-\gamma d(y,y^\prime)^2)$, are no longer guaranteed to be cc-universal. \cite{Christmann10universalkernels} showed that { for compact $\Omega\lo Y$}, if there exists a separable
Hilbert space $\Hcal$ and a continuous injection $\rho : \Omega_Y \to \Hcal$, then for any analytic function $F:\Rbb \to \Rbb$ whose Taylor series at zero has strictly positive coefficients, the function $\kappa(y, y^\prime) = F (\langle\rho(y),\rho(y^\prime)\rangle_\Hcal)$ defines a c-universal kernel on $\Omega_Y$. They also provide an analogous definition of the Gaussian-type kernel in the above case. {  We extend this result to construct cc-universal kernels on non-compact metric space. The proof is given in the Supplementary Material.
\begin{proposition}\label{prop: universal_kernel}
    Suppose $(\Omega_Y, d)$ is a complete and separable metric space, and there exists a separable Hilbert space $\Hcal$ and a continuous injection $\rho : \Omega_Y \to \Hcal$. If $F : \Rbb \to \Rbb$ is an analytic function of the form
    $
        F(t)=\sum_{n=0}^\infty a_n t^n, a_n\ge 0\,\,  \text{for all}\,\, n\ge 1,
    $
    then the function $\kappa : \Omega_Y \times \Omega_Y \to \Rbb$ defined by $\kappa(y,y^\prime)=F(\langle\rho(y),\rho(y^\prime)\rangle_{\Hcal})$ is a positive definite kernel. Furthermore, if $a_n>0$ for all $n\ge 1$, then $\kappa$ is a cc-universal kernel on $\Omega\lo Y$.
\end{proposition}}
As an example, Corollary \ref{coro:gauss} shows that the Gaussian-type kernel is cc-universal on $\Omega_Y$.
\begin{coro}\label{coro:gauss}
Suppose the conditions in Proposition \ref{prop: universal_kernel} are satisfied, then the Gaussian-type kernel
$
    \kappa_\gamma(y,y^\prime)=\exp(-\gamma\|\rho(y)-\rho(y^\prime)\|_{\Hcal}^2),   \ \text{where}\, \ \gamma>0,
$ is cc-universal. Furthermore, if the continuous function $\rho:\Omega_Y\to \Hcal$ is isometric, that is, $d(y,y^\prime)=\|\rho(y)-\rho(y^\prime)\|_{\Hcal}$, then Gaussian-type kernel $\kappa_\gamma(y,y^\prime)=\exp(-\gamma d^2(y,y^\prime))$ is cc-universal.
\end{coro}

The second part of Corollary \ref{coro:gauss} is straightforward since an isometry is an injection. Similar results can  be established for Laplacian-type kernel $\kappa_\gamma(y,y^\prime)=\exp(-\gamma\|\rho(y)-\rho(y^\prime)\|_{\Hcal})$.

The continuous embedding condition in Proposition \ref{prop: universal_kernel} covers several metric spaces often encountered in statistical applications. Section \ref{sec: important case} employs it to construct cc-universal kernels on the space of univariate distributions endowed with Wasserstein-2 distance,  correlation matrices endowed with Frobenius distance, and spheres endowed with geodesic distance.

{
By using the notion of regular probability measure, we connect the cc-universal kernel on $(\Omega_Y,d)$ with the CMS-ensemble, which is the theoretical foundation of our method. Recall that a measure $P\lo Y$ on $(\Omega \lo Y, d )$ is regular if, for any Borel subset $B\subseteq \Omega\lo Y$ and any $\varepsilon>0$, there is a compact set $K\subseteq B$ and an open set $G\supseteq B$, such that $P(G\backslash K)<\varepsilon$.

\begin{theorem}\label{thm: construct}  Suppose, on metric space $(\Omega\lo Y, d)$,
(1) $\ka$ is a bounded cc-universal kernel and (2) $P \lo Y$ is a regular probability measure.
Then the family $\mathfrak F = \{ \ka (\cdot, y): y \in \Omega \lo Y \}$ is a CMS-ensemble.
\end{theorem}

The proof of Theorem \ref{thm: construct} is given in the Supplementary Material. Condition (2), which requires $P\lo Y$ to be regular, is quite mild: it is known that any Borel measure on a complete and separable metric space is regular (see  \citet[Chapter 2: Theorem 1.2, Theorem 3.2]{granirer_1970}). Thus, a sufficient condition of Condition (2) is $(\Omega\lo Y, d)$ being complete and separable, which is satisfied by all the metric spaces we consider. Specifically, note that if $M$ is separable and complete, then so is the Wasserstein-2 space $W\lo2(M)$ \citep[Proposition 2.2.8, Theorem 2.2.7]{panaretos2020invitation}. Therefore, $W\lo 2(\real)$ is complete and separable. Similarly, the SPD matrix space endowed with Frobenius distance and the sphere endowed with geodesic distance are both completely separable metric spaces. Furthermore, the Gaussian kernel and Laplacian kernel we considered satisfy Condition (1) in Theorem 1.}

Thus, Proposition \ref{prop: universal_kernel} and Theorem \ref{thm: construct} provide a general mechanism to construct the CMS-ensemble over any separable and complete metric space without a linear structure, provided it can be continuously embedded in a separable Hilbert space. {  For the case where multiple cc-universal kernels exist, we design a cross-validation framework in Section \ref{sec:simu} to choose the kernel type and  the bandwidth $\gamma$.}


\section{Important Metric Spaces and their CMS Ensembles}\label{sec: important case}
This section gives the construction of CMS-ensembles for three commonly used metric spaces.
\subsection{Wasserstein space}
Let $I$ be $\Rbb$ or a closed interval of $\Rbb$, $\Bcal(I)$ the $\sigma$-field of Borel subsets of $I$, and $\Pcal(I)$ the collection of all probability measures on $(I,\Bcal(I))$. The Wasserstein space $\W$ is defined as the subset of $\Pcal(I)$ with finite second moment, that is,
$
    \W=\{\mu\in \Pcal(I):\int_It^2\, d\mu(t)<\infty\},
$
endowed with the quadratic Wasserstein distance
$
    d_{\mathrm{W}}(\mu_1,\mu_2)=\left(\int_0^1\left[F_{\mu_1}^{-1}(s)-F_{\mu_2}^{-1}(s)\right]^2\,ds\right)^{1/2},
$
where $\mu_1$ and $\mu_2$ are members of $\W$ and $F^{-1}_{\mu_1}$ and $F^{-1}_{\mu_2}$ are the quantile functions of $\mu_1$ and $\mu_2$, which
we assume to be well defined. This distance can be equivalently written as
$
    d_{\mathrm{W}}(\mu_1,\mu_2)=\left(\int_I\left[F_{\mu_1}^{-1}\circ F_{\mu_2}(t)-t\right]^2\,d\mu_2(t)\right)^{1/2}.
$
The set $\W$ endowed with $d_{\mathrm{W}}$ is a metric space with a formal Riemannian structure \citep{ambrosio2004gradient}.

Here, we present some basic results that characterize $\W$, whose proofs can be found, for example, in \cite{ambrosio2004gradient} and \cite{bigot2017geodesic}. For $\mu_1,\,\mu_2\in\W$, we say that a $\Bcal(I)$-measurable map $r: I\to I$ transports $\mu_1$ to $\mu_2$ if $\mu_2=\mu_1\circ r^{-1}$. This relation is often written as $\mu_2={r}_{\#}\mu_1$. Let $\mu_0 \in \W$ be a reference measure with a continuous $F_{\mu_0}$. The tangent space at $\mu_0$ is
$
    T_{\mu_0}=\mathrm{cl}_{L_2(\mu_0)}{\{\lambda(F_\mu^{-1}\circ F_{\mu_0}-\mathrm{id}):\mu\in\W, \lambda > 0\}},
$
where, for a set $A\subseteq L_2(\mu_0)$, $\mathrm{cl}_{L_2(\mu_0)}(A)$ denotes the $L_2(\mu_0)$-closure of $A$. The exponential map $\exp_{\mu_0}$ from $T_{\mu_0}$ to $\W$, defined by $\exp_{\mu_0}({r})=({r}+\mathrm{id})_\#\mu_0$, is surjective. Therefore its inverse, $\log_{\mu_0}: \W\to T_{\mu_0}$, defined by $\log_{\mu_0}(\mu) = F_\mu^{-1} \circ F_{\mu_0}-\mathrm{id}$, is well defined on $\W$. It is well known that the exponential map restricted to the image of $\log$ map, denoted as   $\exp_{\mu_0}|_{\log_{\mu_0}(\mu)(\W)}$, is an isometric homeomorphism \citep{bigot2017geodesic}. Therefore, $\log_{\mu_0}$ is a continuous injection from $\W$ to $L_2(\mu_0)$. We can then construct CMS-ensembles using the general constructive method provided by Theorem \ref{thm: construct} and Proposition \ref{prop: universal_kernel}. The next proposition gives two such constructions, where the subscripts ``G" and ``L" for the two kernels refer to ``Gaussian" and ``Laplacian", respectively.

\begin{proposition}\label{prop:cha_wars}
For $I\subseteq \real$,  $\kappa_G(y,y^{\prime})=\exp(-\gamma\|\log_{\mu_0}(y)-\log_{\mu_0}(y^{\prime})\|^2_{\Lcal_{\mu_0}^2})=\exp(-\gamma d_{\mathrm{W}}(y,y^{\prime})^2)$ and $\kappa_L(y,y^{\prime})=\exp(-\gamma\|\log_{\mu_0}(y)-\log_{\mu_0}(y^{\prime})\|_{\Lcal_{\mu_0}^2})=\exp(-\gamma d_{\mathrm{W}}(y,y^{\prime}))$ are both cc-universal kernels on $\W$. Consequently, the families $\Ffr_G=\{\exp(-\gamma d_{\mathrm{W}}(\cdot,t)^2): t\in \W\}$ and $\Ffr_L=\{\exp(-\gamma d_{\mathrm{W}}(\cdot,t)): t\in \W\}$ are CMS-ensembles.
\end{proposition}

\subsection{Space of symmetric positive definite matrices}
We first introduce some notations. Let $\Sym(r)$ be the set of $r\times r$ invertible symmetric matrices with real entries and $\Sym^+(r)$ the set of $r\times r$ symmetric positive definite (SPD) matrices. For any $Y\in\Rbb\hi{r\times r}$, the \emph{matrix exponential} of $Y$ is defined as the infinite power series $\exp(Y)=\sum_{k=0}^\infty Y^k/k!$. For any $X \in \Sym^+(r)$, the \emph{matrix logarithm} of $X$ is defined as any $r\times r$ matrix $Y$ such that $\exp(Y) = X$ and denoted by $\log(X)$.

Let $d_{\mathrm{F}}$ be the Frobenius metric. Then $(\mathrm{Sym}(r), d_{\mathrm{F}})$ is a metric space continuously embedded by identity mapping in ${\mathrm{Sym}}(r)$, which is a Hilbert space with the Frobenius inner product $\langle A,B\rangle=\tr(A\trans B)$. Also, the identity mapping $\mathrm{id}:\mathrm{Sym}^+(r)\to \mathrm{Sym}(r)$ is obviously isometric. Therefore, by Corollary \ref{coro:gauss}, the two types of radial basis function kernels for Wasserstein space can be similarly extended to $\Sym^+(r)$. That is,  let  $\kappa_G(y,y^{\prime})=\exp(-\gamma d_{\mathrm F}(y,y^{\prime})^2)$ and $\kappa_L(y,y^{\prime})=\exp(-\gamma d_{\mathrm F}(y,y^{\prime})),$ then  $\Ffr_G=\{\kappa_G(y,y^\prime), y^\prime\in \Sym^+(r)\}$ and  $\Ffr_L=\{\kappa_L(y,y^\prime), y^\prime\in \Sym^+(r)\}$ are CMS-ensembles.

Another widely used metric over $\Sym^+(r)$ is the log-Euclidean distance that is defined as
$
    d_{\log}(Y_1,Y_2)=\|\log(Y_1)-\log(Y_2)\|_{\mathrm{F}}.
$
Basically, it pulls the Frobenius metric on $\Sym(r)$ back to $\Sym^+(r)$ by the matrix logarithm map. The matrix logarithm $\log(\cdot)$ is a continuous injection to Hilbert $\Sym(r)$. By Corollary \ref{coro:gauss}, the two types of radial basis function kernels
$\kappa_{G,\log}(y,y^{\prime})=\exp(-\gamma d_{\log}(y,y^{\prime})^2)$ and $\kappa_{L,\log}(y,y^{\prime})=\exp(-\gamma d_{\log}(y,y^{\prime}))$
are cc-universal. Then, $\Ffr_{G,\log}=\{\kappa_{G,\log}(y,y^\prime), y^\prime\in \W\}$ and  $\Ffr_{L,\log}=\{\kappa_{L,\log}(y,y^\prime), y^\prime\in \W\}$ are CMS-ensembles.
\subsection{The sphere}
Consider the random vector taking values in the sphere $\mathbb{S}^n=\{x\in\Rbb^n: \|x\|=1\}$. To respect the nonzero curvature of $\mathbb{S}^n$, the geodesic distance $d_g(Y_1,Y_2)=\arccos(Y_1\trans Y_2)$, which is derived from its Riemannian geometry, is often used rather than the Euclidean distance. However, the popular Gaussian-type RBF kernel $\kappa_G(y,y^{\prime})=\exp(-\gamma d_g(y,y^\prime)^2)$ is not positive definite on $\mathbb{S}^n$ \citep{jayasumana2013combining}. In fact, \cite{feragen2015geodesic} proved that for complete Riemannian manifold $M$ with its associated geodesic distance $d_g$,  $\kappa_G(y,y^{\prime})=\exp(-\gamma d_g(y,y^\prime)^2)$ is positive semidefinite only if $M$ is isometric to a Euclidean space. \cite{honeine2010angular} and \cite{jayasumana2013combining} proved that the Laplacian-type kernel $\kappa_L(y,y^{\prime})=\exp(-\gamma d_g(y,y^\prime))$ is positive definite on the sphere $\mathbb{S}^n$. We show in the following proposition that $\kappa_L(y,y^{\prime})$ is cc-universal.
\begin{proposition}\label{prop:sphere_universal}
The Laplacian-type kernel $\kappa_L(y, y^\prime) : \mathbb{S}^n\times
\mathbb{S}^n\to \Rbb$, defined by $\kappa_L(y, y^\prime) = \exp(-\gamma d_g(y,y^\prime))$, where $d_g$ is the geodesic distance on $\mathbb{S}^n$, is a cc-universal kernel for any $\gamma>0$. Consequently, $\Ffr_L=\{\exp(-\gamma d_g(\cdot,t)), t\in \mathbb{S}^n\}$ is a CMS-ensemble.
\end{proposition}

\section{Fr\'echet Sufficient Dimension Reduction}\label{sec:fsdr}
In this section, we develop the Fr\'echet SDR estimators based on the CMS-ensembles and CS-ensembles and establish their Fisher consistency.

\subsection{Ensembled moment estimators via CMS ensembles}\label{subsection: ensemeble cms moments}
We first develop a general class of Fr\'echet SDR estimators based on the ensembled moment estimators of the CMS, such as the OLS, PHD, and IHT. Let $\ca P\lo {XY}$ be the collection of all distributions of $(X,Y)$, and let $M: \ca P\lo {XY} \to \real \hi {p \times p}$ be a measurable function to be used as an estimator of the Fr\'echet central subspace $\ca S \lo {Y|X}$. A function defined on $\ca P\lo {XY}$ is called statistical functional; see, for example, Chapter 9 of \cite{li2018sufficient}. In the SDR literature, such a function is also called a candidate matrix \citep{ye2003using}. Let $F \lo {XY}$ be a generic member of $\ca P\lo {XY}$, $F \lo {XY} \hii 0$ the true distribution of $(X,Y)$, and $\hat F \lo {XY} \hii n$ the empirical distribution of $(X,Y)$ based on an i.i.d. sample $(X \lo 1, Y \lo 1), \ldots, (X \lo n , Y \lo n)$. Extending the terminology of classical SDR (see, for example, \citealt{li2018sufficient}, Chapter 2), we say that the estimate $M ( \hat F \lo {XY} \hii n)$ is unbiased if $M( F \lo {XY} \hii 0 ) \subseteq \ca S \lo {Y|X}$, exhaustive if  $M( F \lo {XY} \hii 0 ) \supseteq \ca S \lo {Y|X}$, and Fisher consistent if $M( F \lo {XY} \hii 0 ) = \ca S \lo {Y|X}$. We refer to $M$ as the Fr\'echet candidate matrix.

Suppose we are given a CMS-ensemble $\mathfrak F$. Let $M \lo 0: \ca P\lo {XY} \times \mathfrak F \to \real \hi {p \times p}$ be a function to be used as an estimator of $\ca S \lo {E[f(Y)|X] }$ for each $f$. This is not a statistical functional in the classical sense, as it involves an additional set $\mathfrak F$. So, we redefine unbiasedness, exhaustiveness, and Fisher consistency for this type of augmented statistical functional.

\begin{definition}\label{definition:new Fisher consistency} We say that $M \lo 0$ is unbiased  for estimating $\{ \ca S \lo {E[f(Y)|X]}: f \in \mathfrak F \}$ if, for each $f \in \mathfrak F$,
$
\spn \{M \lo 0 ( F \lo {XY} \hii 0, f) \} \subseteq  \ca S \lo {E[f(Y)|X]}.
$
Exhaustiveness and Fisher consistency of $M \lo 0$ are defined by replacing $\subseteq$ in the above by $\supseteq$ and $=$, respectively.
\end{definition}
Note that $M \lo 0 (\cdot, f)$ is an estimator of the classical central mean subspace $\ca S \lo {E[f(Y)|X]}$, as $f(Y)$ is a random number rather than a random object. We refer to  $M \lo 0$ as the ensemble candidate matrix or, when confusion is possible, the CMS-ensemble candidate matrix. Our goal is to construct a Fr\'echet candidate matrix $M: \ca P\lo {XY} \to \real \hi {p \times p}$ from the ensemble candidate matrix $M \lo 0: \ca P\lo {XY} \times \mathfrak F \to \real \hi {p \times p}$. To do so, we assume $\mathfrak F$ is of the form $\{ \ka (\cdot, y ): y \in \Omega \lo Y \}$, where $\ka: \Omega \lo Y \times \Omega \lo Y \to \real $ is a cc-universal kernel. Given such an $\mathfrak F$ and $M \lo 0$, we define $M$ as follows
\begin{align*}
M ( F \lo {XY} ) = \int \lo {\Omega \lo Y} M \lo 0 ( F \lo {XY}, \ka (\cdot, y) ) d F \lo Y (y),
\end{align*}
where $F \lo Y$ is the distribution of $Y$ derived from $F \lo {XY}$.

We now adapt several estimates for the classical central mean subspace to the estimation of Fr\'echet SDR: the ordinary least squares (OLS; \citealt{li1989regression}), the principal Hessian directions (PHD; \citealt{li1992principal}), and the Iterative Hessian Transformation (IHT; \citealt{cook2002dimension}). These estimates are based on sample moments and require additional conditions on the predictor $X$ for their unbiasedness. Specifically, we make the following assumptions :
\begin{assumption}\label{assumption: lcm and ccv}
1.Linear Conditional Mean (LCM): $E(X|\beta\trans X)$ is a linear function of $\beta\hi\trans X$, where $\beta$ is a basis matrix of the Fr\'echet central subspace $\ca S \lo {Y|X}$; \\
2. Constant Conditional Variance (CCV): $\var(X | \beta\trans X)$ is a nonrandom matrix.
\end{assumption}
Under the first assumption, the  ensemble OLS and IHT are unbiased for estimating the Fr\'echet central subspace; under both assumptions, the ensemble PHD is unbiased for estimating $\ca S \lo {Y|X}$. More detailed discussions on the unbiasedness and fisher consistency of ensemble estimators are presented in Section \ref{subsection: fisher consistency}. { In practice, the two assumptions above cannot be checked directly since we do not know $\beta$. However, as was shown by \citet{eaton1986characterization}, if Assumption 1 holds for all $\beta$, then the distribution of $X$ is elliptical, and vice versa. If further $X$ is multivariate normal, then Assumption 2 is satisfied. Currently, the scatter plot matrix is the most commonly used empirical method to check the elliptical distribution assumption. If non-elliptical features are observed, one can use marginal transformations of the predictors, such as the Box-Cox transformation, to mitigate the non-ellipticity problem. Furthermore, in practice, the SDR methods that require ellipticity usually still work reasonably well even when the elliptical distribution assumption is violated. This occurs particularly when the dimension $p$ of $X$ is high. See \cite{hall1993almost} and \citet{li2007surrogate} for the theoretical supports. Our simulation results in Section \ref{sec:simu} support this phenomenon. }

It is most convenient to construct these ensemble estimators using standardized predictors. The theoretical basis for doing so is an equivariant property of the Fr\'echet central subspace, as stated in the next proposition.

\begin{proposition} If $\ca S \lo {Y|X}$ is the Fr\'echet central subspace, $A \in \real \hi {p \times p}$ is a non-singular matrix, and $b$ is a vector in $\real \hi p$, then
$
\ca S \lo {Y|A X + b} = A \trans S \lo {Y|X}.
$
\end{proposition}
The proof is essentially the same as that for the classical central subspace (see, for example, \citealp{li2018sufficient}, page 24), and is omitted. Using this property, we first transform $X$ to
$Z = \var(X) \hi {-1/2} (X - EX)$, estimate the Fr\'echet central subspace $\ca S \lo {Y|Z}$, and then transform it by $\var (X) \hi {-1/2} \ca S \lo {Y|Z}$, which is the same as $\ca S \lo {Y|X}$.
The candidate matrices $M \lo 0$ and $M$ for ensemble OLS, PHD, and IHT are formulated in Remark \ref{remark: candidate matrix}. Detailed motivation for each can be found in \citet[Chapter 8]{li2018sufficient}. The sample estimates can then be constructed by replacing the expectations in $M \lo 0$ and $M$ with sample moments whenever possible. Algorithm \ref{algo: FOLS} summarize the steps to implement an ensembled moment estimator, where $\ka \lo c (y,y')$ stands for the centered kernel $\ka (y,y') - E \lo n \ka (Y, y')$.

\smallskip
\smallskip
\smallskip

{
\begin{algorithm}[H]
\SetAlgoLined
\justify
\textbf{Step 1}. Standardize predictors. Compute sample mean $\hat\mu=E_n(X)$ and sample variance $\hat{\Sigma}=\var_n(X)$. Then let
$ Z_i=\hat\Sigma^{-1/2}(X_i-\hat\mu)$.\\
\textbf{Step 2}. Compute $\hat{M}_0 (y)$ for $y=Y_1,\dots,Y_n$, with specific form given in Remark \ref{remark: candidate matrix}.\\
\textbf{Step 3}. Compute $\hat M=\frac{1}{n}\sum_{i=1}^n \hat{M}_0(Y_i)$.\\
\textbf{Step 4}. Let $\hat v_1,\dots, \hat v_{d_0}$ be the leading $d_0$ eigenvectors of $\hat M$, and let $u_k = \hat\Sigma^{-1/2}v_k$, for $k = 1,\dots,d_0$. Then use $\{u_1,\dots,u_{d_0}\}$ to estimate a basis of the Fr\'echet central subspace $\Scal_{Y|X}$.
 \caption{Fr\'echet OLS, PHD, IHT, SIR, SAVE, and DR}\label{algo: FOLS}
\end{algorithm}
\begin{remark}\label{remark: candidate matrix}
The candidate matrix $M\lo0(y)$ for Fr\'echet OLS, PHD, and IHT are
\begin{itemize}
    \item[(1)] (Fr\'echet OLS) ${M}_0 (y)= {C}(y){C}(y) \trans$, where ${C}(y) = \cov[Z, \ka (Y, y)]$;
    \vspace{-0.1in}
    \item[(2)] (Fr\'echet PHD) ${M}_0 (y)= E[Z Z \trans\kappa \lo c(Y, y)]$;
    \vspace{-0.1in}
    \item[(3)] (Fr\'echet IHT)  ${M}_0 (y)= {W}(y)W(y)$, where  ${H}(y)=E[{ZZ\trans \kappa\lo c(Y,y)}], {W}(y) = ({C}(y), {H}(y) {C}(y), \ldots, {H} (y) \hi r {C} (y) )$.
\end{itemize}
\end{remark}
}
\subsection{Ensembled forward regression estimators via CMS ensembles}

In this subsection, we adapt the OPG (\citealt{xia2002adaptive}),  a popular method for estimating the classical CMS based on nonparametric forward regression, to the estimation of the Fr\'echet central subspace, which do not require LCM and CCV conditions. The adaption of another forward regression method MAVE is similar and presented in Section S.3.2 of the Supplementary Material.  The framework of the statistical functional $M\lo 0(F\lo{XY}, f)$ is no longer sufficient to cover this case because we now have a tuning parameter here. So, we adopt the notion of tuned statistical functional in Section 11.2 of \cite{li2018sufficient} to accommodate a tuning parameter.

Let $\ca P\lo {XY}$, $F \lo {XY}$, $F \lo {XY} \hi {(0)}$ and $\hat F \lo {XY} \hii n$ be as defined in Section \ref{subsection: ensemeble cms moments}. For simplicity, we assume the tuning parameter $h$ to be a scalar, but it could also be a vector. Given a CMS-ensemble $\Ffr$, let $T_0: \ca P\lo {XY}\times \Ffr\times \Rbb\to \Rbb\hi {p\times p}$ be a tuned functional to be used as an estimator of $\ca S \lo {E[f(Y)|X]}$ for each $f$. We refer to $T_0$ as the ensemble-tuned candidate matrix. The unbiasedness, exhaustiveness, and Fisher consistency of $T_0$ are defined as follows.
\begin{definition}\label{definition:new tuned Fisher consistency} We say that $T \lo 0$ is unbiased  for estimating $\{ \ca S \lo {E[f(Y)|X]}: f \in \mathfrak F \}$ if, for each $f \in \mathfrak F$,
$
\spn \{\lim _{h\to 0} T \lo 0 (F \lo {XY} \hii 0, f, h) \} \subseteq  \ca S \lo {E[f(Y)|X]}.
$
Exhaustiveness and Fisher consistency of $T \lo 0$ are defined by replacing $\subseteq$ in the above by $\supseteq$ and $=$, respectively.
\end{definition}
Given $\Ffr = \{\kappa(\cdot, y): y\in\Omega_Y\}$ and $T\lo 0$, we define the tuned Fr\'echet candidate matrix $T: \ca P \lo {XY} \times \Rbb \to \Rbb^{p\times p}$
as $T(F\lo {XY}, h) = \int_{\Omega_Y}T_0(F\lo {XY}, \kappa(\cdot, y), h)d F\lo Y (y).$
We say that the estimate $T(T\hi{(n)}\lo{XY}, h)$ is unbiased if $\spn(\lim\lo {h\to 0}T(F\hi {(0)}\lo {XY}, h))\subseteq \ca S\lo{Y|X}$, exhaustive if $\spn(\lim\lo {h\to 0}T(F\hi {(0)}\lo {XY}, h))\supseteq \ca S \lo {Y|X}$, and Fisher consistent if $\spn(\lim\lo {h\to 0}T(F\hi {(0)}\lo {XY}, h)) =\ca S\lo {Y|X}$.

For a function $h(x)$, we use $\partial h(X) / \partial X$ to denote $\partial h(x) / \partial x $ evaluated at $x = X$. The OPG aims to estimate central mean subspace $\ca S \lo {E[\kappa(Y, y)|X]}$ by
$
E \left[
\frac{ \partial E ( \ka (Y, y) | X  ) }{\partial X} \frac{ \partial E ( \ka (Y, y) | X  ) }{\partial X \trans} \right]
$
where the gradient $\partial E ( \ka (Y, y) | X  )/{\partial X}$ is estimated by local linear approximation as follows. Let $K_0:\Rbb \to [0,\infty)$ be a kernel function as used in kernel estimation. For any $v \in \Rbb^p$ and bandwidth $h>0$, let $K_h(v) = h^{-p}K_0(\|v\|/h)$. At the population level, for fixed $x\in\Omega_X$ and $y\in\Omega_Y$, we minimize the objective function
\begin{equation}\label{eq:obj_pop_opg}
    E\{[\kappa(Y,y) - a - b\trans(X -x)]^2K_h(X - x)\}/EK_h(X - x)
\end{equation}
over all $a\in\Rbb$ and $b\in \Rbb^{d_0}$.
The minimizer depends on $x, y$ and we write it as $(a_h(x,y)$, $b_h(x,y))$. The ensemble tuned candidate matrix for estimating the central mean subspace $\ca S \lo {E[\kappa(Y, y)|X]}$ is $T\lo 0 (F\lo {XY}, \kappa(\cdot, y), h) = E[b\lo h(X, y)b\lo h(X, y)\trans]$ and the tuned Fr\'echet candidate matrix is $T(F\lo {XY}, h) = E[b\lo h(X, Y)b\lo h(X, Y)\trans]$.

At the sample level, we minimize, for each $j,k=1,\dots,n$, the empirical objective function
\begin{equation}\label{eq:obj_emp_opg}
    \sum_{i=1}^nw_{h}(X_i,X_j)\left[\kappa_\gamma(Y_i,Y_k)-a_{jk}-b_{jk}\trans(X_i-X_j)\right]^2
\end{equation}
over $a_{jk}\in \Rbb$ and $b_{jk}\in\Rbb^{p}$, where
$w_{h}(X_i,X_j)=K_h(X_i-X_j)/\sum_{l=1}^nK_h(X_l-X_j).$
Following \cite{xia2002adaptive}, we take the bandwidth to be $h=c_0n^{1/(p_0+6)}$ where $p_0=\max\{p,3\}$ and $c_0=2.34$, which is slightly larger than the optimal $n^{-1/(p+4)}$ in terms of the mean integrated squared errors. As proposed in \citet[Lemma 11.6]{li2018sufficient}, instead of solving $b_{jk}$ from (\ref{eq:obj_emp_opg}) $n^2$ times, we solve multivariate
weighted least squares to obtain $b_{j1},\dots,b_{jn}$ simultaneously. { Computation details are given in Section S.3 of the Supplementary Material.}

We can further enhance the performance of FOPG by projecting the original predictors onto the directions produced by the FOPG to re-estimate $\Scal_{Y|X}$. Specifically, after the first round of FOPG, we form the matrix
$\hat{B} = (\hat{v}_1, \dots ,\hat{v}_d)$ and replace the
kernel $K_h(X_j-X_i)$ by $K_h(\hat B\trans(X_j-X_i))$ with an updated bandwidth $h$, and complete the next round of iteration, which leads to an updated $\hat B$. We then iterate this process until convergence. In this way, we reduce the dimension of the kernel from $p$ to $d_0$ and mitigate the “curse of dimensionality”.  To avoid confusion, We call this refined version of Fr\'echet OPG as FOPG in the following. The algorithms for FOPG are summarized as Algorithm \ref{algo: fopg}.

\begin{algorithm}[ht]
\SetAlgoLined
\justify
\textbf{Step 1}. Standardize $X_1,\dots ,X_n$ by
$Z_a^i = (X_a^i-\bar X^i)/\hat\sigma_i$,
where $X^i_a$ denotes the $i$-th component of $X_a$ and $\hat\sigma_i=\sqrt{\var_n(X^i)}$. Set $\hat{B}^{(0)} = I_p$ as the initial value and set the maximum number of iterations as 10. Set iteration time $t$ to 1.\\
\textbf{Step 2}. For each $j=1, \dots, n$ and iteration $t$, solving vectors $\hat b^{(t)}_{j1},\dots,\hat b^{(t)}_{jn}$ from \eqref{eq:obj_emp_opg}.\\
\textbf{Step 3}. Compute
$
    \hat\Lambda^{(t)}=\frac{1}{n^2}\sum_{j,k=1}^n\hat b^{(t)}_{jk}\left(\hat b^{(t)}_{jk}\right)\trans.
$
Then perform eigen-decomposition for $\hat\Lambda^{(t)}$ and get the $d_0$ eigenvectors corresponding to its largest eigenvalues $\hat{v}_1,\cdots, \hat{v}_{d_0}$. Let $\hat{B}_{(t)}=(\hat{v}_1,\cdots, \hat{v}_{d_0})$.\\
\textbf{Step 4}. If $t\le 10$, reset $ h_t = \max(r_nh_{t-1}; c_0n^{-1/(d+4)})$, where $r_n=n^{-1/2(p_0+6)}$, increase $t$ by 1 and return to step 2. Otherwise, Let $\hat D$ be the diagonal matrix $\diag(\hat\sigma_1,\dots,\hat\sigma_p)$. A basis of the central subspace $S_{Y|X}$ is estimated by $\{\hat D^{-1/2}\hat{v}_1,\dots,\hat D^{-1/2}\hat{v}_{d_0}\}$.
\caption{Fr\'echet OPG}\label{algo: fopg}
\end{algorithm}

\subsection{Ensembled inverse regression estimators via CS ensembles}\label{subsection: ensemble cs}

In this subsection, we adapt several well-known estimators for the classical central subspace to Fr\'echet SDR, which include SIR \citep{li1991sliced}, SAVE \citep{cook1991sliced}, and DR \citep{li2007directional}. We use the CS-ensemble to combine these classical estimates through  \eqref{eq:csens}.
Let $\mathfrak{F} = \{ \ka (\cdot, y): y \in \Omega \lo Y\}$ be a CS ensemble, where $\ka$ is a cc-universal kernel. Let $M \lo 0:
\ca P\lo {XY} \times \mathfrak F \to \real \hi {p \times p}$ be a CS-ensemble candidate matrix. Let $M(F \lo {XY}) = \int M \lo 0 (F \lo {XY}, \ka (\cdot, y)) d F \lo Y (y)$ be the Fr\'echet candidate matrix.

Again, we work with the standard predictor $Z$. The candidate matrices $M \lo 0(y)$ for ensemble SIR, SAVE, and DR are formulated in Remark \ref{remark: candidate matrix 2}. Detailed motivation for each can be found in \citet[Chapter 3,5,6]{li2018sufficient}. At the sample level, we replace any unconditional moment $E$ by the sample average $E \lo n$, and replace any conditional moment, such as $E(Z|\kappa(Y,y))$, by the slice mean. The algorithms are also included in Algorithm \ref{algo: FOLS}. A more detailed algorithm is given by Algorithm 5 in the Supplementary Material.
\begin{remark}\label{remark: candidate matrix 2}
The candidate matrices $M \lo 0 (y)$ for Fr\'echet SIR,  SAVE, and  DR are $  \var [ E (Z | \ka (Y, y)]$, $ [ I \lo p - \var (Z | \ka (Y, y))] \hi 2 $, and
$ 2 E \{ E [ Z Z \trans | \ka (Y, y) ]\} \hi 2 + 2 E \hi 2 \{ E[ Z | \ka (Y, y)] E [ Z \trans | \ka (Y, y)] \}  + 2 E \{ E [ Z \trans | \ka (Y,y)] E [ Z  | \ka (Y,y)] \} \, E \{ E [ Z | \ka (Y,y)] E [ Z  \trans | \ka (Y,y)] \} - 2 I \lo p$, respectively.
\end{remark}
{
\begin{remark}
Regarding the time complexity of the Frechet SDR methods, by construction, the ensemble estimator requires $n$ times the computing time of the original estimator because it needs to reapply the original estimator for each $\ka (\cdot, y \lo i)$, $i=1, \ldots, n$. For example, if SAVE is used as the original estimator, then the largest matrix multiplication is $A \lo {p \times n} B \lo {n \times p}$ which requires  $ p \hi 2 n$ basic computation units; the largest matrix to invert or eigen decomposition to perform is a $p \times p$ matrix, which requires $p \hi 3$ basic computation units. So the net computation complexity is $n \times  \max( O(n p \hi 2), O (p \hi 3))$.
\end{remark}}

\subsection{Fisher consistency}\label{subsection: fisher consistency}

In this subsection, we establish the unbiasedness and Fisher consistency of the tuned Fr\'echet candidate matrix. As a special case, the Fr\'echet candidate matrix constructed by any moment-based methods in Section \ref{subsection: ensemeble cms moments} can be considered as tuned Fr\'echet candidate matrix with the tuning parameter $h$ taken to be $0$. The next theorem shows that if $T \lo 0$ is unbiased (or Fisher consistent), then $T$ is unbiased (or Fisher consistent). In the following, we say that a measure $\mu$ on $\Omega_Y$ is strictly positive if and only if for any nonempty open set $U\subseteq\Omega_Y$, $\mu(U)>0$. For a matrix $A$, $\|A\|$ represents the operator norm.

\begin{theorem}\label{theorem:Fisher consistency:CMS} Suppose $\mathfrak  F = \{  \ka (\cdot, y): y \in \Omega \lo Y \}$ is a CMS-ensemble, where $\ka$ is a cc-universal kernel. We have the following results regarding unbiasedness and Fisher consistency for $T$.
\begin{enumerate}
\item If $T \lo 0$ is unbiased for $\{ \ca S \lo {E [\ka (Y, y)|X]}: f \in \mathfrak F\}$ and $\|T \lo 0 ( F \lo {XY} \hii 0, \ka (\cdot, Y'), h)\|\le G(Y')$, where $G(Y')$ is a real-valued function with $E[G(Y')]<\infty$, then $T$ is unbiased for  $\ca S \lo {Y|X}$;
\item
If (a) $T \lo 0$ is Fisher consistent for  $\{ \ca S \lo {E [\ka (Y, y)|X]}: f \in \mathfrak F\}$,
(b) $T \lo 0 (F \lo {XY}, \ka ( \cdot, y), h )$ is positive semidefinite for each $y \in \Omega \lo Y$, $h\in \Rbb$ and $F \lo {XY} \in \ca P\lo {XY}$, (c) $\lim{ \sup}\lo{h\to 0}\|T \lo 0 ( F \lo {XY} \hii 0, \ka (\cdot, Y'), h)\|\le G(Y')$ with $E[G(Y')]<\infty$, (d) $F_Y$ is strictly positive on $\Omega_Y$, and (e) the mapping $y' \mapsto \lim\lo{h\to 0}T \lo 0 ( F \lo {XY}, \ka (\cdot, y'), h)$ is continuous, then
$T$ is  Fisher consistent  for  $\ca S \lo {Y|X}$.
\end{enumerate}
\end{theorem}
We similarly develop Fisher consistency for Fr\'echet SDR based on the CS-ensemble, including methods in Section 4.3. The next corollary says that if $M \lo 0$ is Fr\'echet consistent for $\{ \ca S \lo {\ka (Y,y)|X}: y \in \Omega \lo Y\}$, then $M$ is Fr\'echet consistent for $\ca S \lo {Y|X}$. The proof is similar to that of Theorem \ref{theorem:Fisher consistency:CMS} and is omitted.

\begin{coro} Suppose $\mathfrak  F = \{  \ka (\cdot, y): y \in \Omega \lo Y \}$ is a CS-ensemble, where $\ka$ is a cc-universal kernel. We have the following results regarding unbiasedness and Fisher consistency for $M$.
\begin{enumerate}
\item If $M \lo 0$ is unbiased  for  $\{ \ca S \lo {\ka (Y, y)|X }: f \in \mathfrak F\}$, then $M$ is unbiased for  $\ca S \lo {Y|X}$;
\item
If $M \lo 0$ is Fisher consistent   for  $\{ \ca S \lo { \ka (Y, y)|X }: f \in \mathfrak F\}$,
$M \lo 0 (F \lo {XY}, \ka ( \cdot, y) )$ is positive semidefinite for each $y \in \Omega \lo X$ and $F \lo {XY} \in \ca P\lo {XY}$, $F_Y$ is strictly positive, and the mapping $y' \mapsto M \lo 0 ( F \lo {XY}, \ka (\cdot, y'))$ is continuous, then
$M$ is  Fisher consistent  for  $\ca S \lo {Y|X}$.
\end{enumerate}
\end{coro}

{
Unbiasedness and Fisher consistency of $T \lo 0$ or $M\lo 0$ are satisfied by different sets of sufficient conditions for the moment-based or forward-regression-based estimators. We outline these conditions below.
\begin{enumerate}
\item For ensembled moment estimators in Section 4.1 and ensembled inverse regression estimators in Section 4.3, most of them are unbiased under either the LCM assumption or both the LCM and CCV assumption. For example, the unbiasedness of SIR, OLS, and IHT requires the LCM assumption, whereas the unbiasedness of SAVE, DR, and PHD requires both the LCM and the CCV assumptions. The estimators SIR, OLS,  IHT, and PHD  are generally not exhaustive (recall that unbiased along with exhaustiveness is equivalent to Fisher consistency). But sufficient conditions for SAVE, DR to be exhaustive are reasonably mild (see \citet{li2007directional} and \citet[Chapter 6]{li2018sufficient}).

\item Sufficient conditions for Fisher consistency for OPG are given in \citet[Section 11.2]{li2018sufficient}. Specifically, it requires: (a) the smooth kernel function $K\lo 0$ is a spherically-contoured p.d.f. with finite fourth moments; (b)  the p.d.f. of $X$ is supported on $\real\hi p$ and has continuous bounded second derivatives. Note that neither LCM nor CCV assumption is needed for the OPG estimator.
\end{enumerate}}

\section{Convergence Rates of the Ensemble Estimates}\label{sec:asymptotics}
In this section, we develop the convergence rates of the ensemble estimates for Fr\'echet SDR. To save space, we will only consider the CMS-ensemble; the results for the CS-ensemble are largely parallel. To simplify the asymptotic development, we make a slight modification of the ensemble estimator, which does not result in any significant numerical difference from the original ensembles developed in the previous sections. For each $i = 1, \ldots, n$, let  $\hat F \lo {XY} \hi {(-i)}$ be the empirical distribution based on the sample with $i$th subject removed:  $\{(X \lo 1, Y \lo 1), \ldots, (X \lo n, Y \lo n) \} \setminus \{(X \lo i, Y \lo i) \}$. Our modified ensemble estimate is of the form
\begin{align*}
T (\hat F \lo {XY} \hii n, h_n) = n \inv \tsum \lo {i=1} \hi n T \lo 0 ( \hat F \lo {XY} \hi {(-i)}, \ka (\cdot, Y \lo i ), h_n ).
\end{align*}
The purpose of this modification is to break the dependence between the ensemble member $\ka (\cdot, Y \lo i)$ and the CMS estimate, which substantially simplifies the asymptotic argument. Here, we let the tuning parameter $h_n$ depend on $n$. Again, the Fr\'eceht candidate matrix constructed by moment-based methods can be considered as a special case with $h_n = 0$.

Rather than deriving the convergence rate of each individual ensemble estimate case by case, we will show that, under some mild conditions, the ensemble convergence rate is the same as the corresponding CMS-estimate's rate. Since the convergence rates of many CMS-estimates are well established, including all the forward regression and sample moment-based estimates mentioned earlier, our general result covers all the CMS-ensemble estimates.

In this following, for a matrix $A$, $\| A \|$ represents the operator norm and $\| A \| \lo {\mathrm{F}}$ the Frobenius norm. If $\{a \lo n \}$ and $\{b \lo n \}$ are sequences of positive numbers, we write $a \lo n \prec b \lo n$ if $\lim \lo {n \to \infty} a \lo n / b \lo n = 0$; we write $a \lo n \preceq b \lo n$ if $a \lo n /b \lo n$ is a bounded sequence. We write $b \lo n \succ a \lo n$ (or $b \lo n \succeq a \lo n$) if $a \lo n \prec b \lo n$ (or $a \lo n \preceq b \lo n$). We write $a \lo n \asymp b \lo n$ if $a \lo n \preceq b \lo n$ and $b \lo n \preceq a \lo n$. Let $T \lo 0 \hi * (F \lo {XY} \hii {0}, \ka (\cdot, y)) = \lim\lo{h\to 0}T \lo 0 (F \lo {XY} \hii {0}, \ka (\cdot, y), h)$ and $T\hi * (F \lo {XY} \hii 0) = \lim\lo {h\to 0} T(F \lo {XY} \hii 0, h)$.

\setcounter{theorem}{2}
\begin{theorem} Let  $C \lo n (y) = E \| T \lo 0 (\hat F \lo {XY} \hii {n}, \ka (\cdot, y), h\lo n )  - T \lo 0\hi * (F \lo {XY} \hii {0}, \ka (\cdot, y)) \| $ and  $\{ a \lo n \}$ be a positive sequence of numbers satisfying $a \lo {n+1}/a \lo n \asymp 1$ and $a \lo n \succeq n \hi {-1/2}$. Suppose
the entries of $T \lo 0 \hi * (F \lo {XY} \hii 0, \kappa(\cdot, Y) )$ have finite variances. If
$E [C \lo n (Y)] = O (a \lo n)$,  then
$\| T  ( \hat F \lo {XY} \hii n, h\lo n ) - T\hi * (F \lo {XY} \hii 0) \| = O \lo P ( a \lo n)$.
\end{theorem}

The above theorem says that, under some conditions, the convergence rate of an ensemble Fr\'echet SDR estimator is the same as the corresponding CMS estimator. This covers all the estimators developed in Section 4. Specifically:
\begin{enumerate}
    \item For all moment-based ensemble methods, such as OLS, PHD, IHT, SIR, SAVE, DR, the ensemble candidate matrices can be written in the form $M_0(\hat F \lo {XY} \hii n, \kappa(\cdot, y))= \hat \Lambda(y)\hat\Lambda(y)\trans$, where $\hat\Lambda(y)$ is a matrix  possessing the second order von Mises expansion, implying $E[C_n(Y)] = O(n\hi {-1/2})$. See, for example, \cite{li2018sufficient}
    \item For nonparametric forward regression ensemble methods, OPG and MAVE, the convergence rate of $C_n(y)$ was reported in \cite{xia2007constructive} as $O(h\hi 2\lo n + h\lo n\hi{-1}\delta\hi 2\lo n)$ where $\delta\lo n = \sqrt{(\log n)/n h\hi p\lo n}$. Although the convergence was established in terms of convergence in probability, under mild conditions such as uniformly integrability, we can obtain the same rate for $E[C_n(Y)]$.
\end{enumerate}

\section{Simulations}\label{sec:simu}
We evaluate the performance of the proposed Fr\'echet SDR methods with distributions and symmetric positive definite matrices as responses. For space consideration, the additional simulation for spherical data is presented in the Supplementary Material.
{
\subsection{Computational Details}

\textbf{Choice of tuning parameters and kernel types.}
We first implement a unified cross-validation procedure to select the kernel type and bandwidth $\gamma$ in the kernel. For both  distributional response and symmetric positive definite matrix response, we consider  Gaussian radial basis kernel $\kappa\lo {\mathrm G}(y,y') = \exp(- \gamma d(y,y')^2)$ and Laplacian radial basis kernel $\kappa\lo {\mathrm L}(y, y')$ as candidates to construct the ensembles. For the bandwidth $\gamma$, we set the default value as
\begin{equation}\label{eq: kernel_para}
        \gamma\lo {\mathrm G}=\frac{\rho_Y}{2\sigma\lo {\mathrm G}^2},\quad \mbox{where}\quad  \sigma\lo {\mathrm G}^2=\binom{n}{2}\inv\sum_{i<j}d(Y_i,Y_j)^2,\quad \rho_Y=1,
\end{equation}
in the Gaussian radial basis kernel, and
\begin{align*}
        \gamma\lo {\mathrm L}=\frac{\rho_Y}{2\sigma_L},\quad \mbox{where}\quad  \sigma\lo {\mathrm L}=\binom{n}{2}\inv\sum_{i<j}d(Y_i,Y_j),\quad \rho_Y=1,
\end{align*}
in the Laplacian radial basis kernel. The same choices were used in \citet{lee2013general} and \citet{li2017nonlinear}. We then fine-tune $\rho\lo Y$ and kernel types together via the $k$-fold cross-validation as follows. Randomly split the whole sample into $k$ subsets of roughly equal sizes, say $D\lo 1, \dots, D\lo k$. For each $i = 1,\dots, k$, use $D\lo i$ as the test set and its complement as the training set. We first use the training set to implement the Fr\'echet SDR with an initial dimension $d$, say $5$. This choice of a relatively large dimension helps to guarantee the unbiasedness of the estimated Fr\'echet central subspace. We then substitute the estimated $\hat\beta$ into the testing set to produce the sufficient predictor $\hat\beta\trans X$ and then fit a global Fr\'echet regression model \citep{petersen2019frechet} to predict the response in the testing set. Compute the prediction error for each $i$ and aggregate the error for all rotations $i = 1,\dots, k$, which yields an overall cross-validation error. This overall error depends on the tuning parameter $\rho\lo Y$ and kernel type and is then minimized over a grid $\{10\hi {-2}, 10\hi{-1}, 1, 10\}\times\{\kappa\lo {\mathrm{G}}, \kappa\lo{\mathrm{L}}\}$ to obtain the optimal combinations.

\textbf{Estimation of the dimensions.}
For the ensemble estimators that possess a candidate matrix (such as the ensemble moment estimators in Section 4.1), the recently developed order-determination methods, such as the ladle estimate \citep{luo2016combining}, and predictor augmentation estimator \citep{luo2021order} can be directly applied to estimate $d \lo 0$. In addition, the BIC-criterion introduced by \citet{zhu2006sliced} can also be used for this purpose.

In this paper, we adapted the predictor augmentation estimator to the current setting. A detailed introduction of the predictor augmentation method and more simulation results are included in the Supplementary Material. For the predictor augmentation estimator, we take the times of augmentations $s=10$ and the dimension of augmented predictors $r =\lceil p/2 \rceil$, where $p$ is the original dimension of predictors.

\textbf{Estimation error assessment.} We used the error measurement for subspace estimation as in \cite{li2005contour}: if $\ca S\lo 1$ and $\ca S\lo 2$ are two subspaces of $\mathbb{R}\hi p$ of the same dimension, then their distance is defined as
    $
        d(\ca S\lo 1, \ca S\lo 2) = \|P\lo {\ca S\lo 1}-P\lo {\ca S\lo 2}\|\lo{\mathrm{F}},
   $
where $P\lo{\ca S}$ is the projection on to $\ca S$, and $\|\cdot\|\lo{\mathrm{F}}$ is the Frobenius matrix norm. If $B \lo 1$ and $B \lo 2$ are two matrices whose columns form bases of $\ca S \lo 1$ and $\ca S \lo 2$ respectively, this distance can be equivalently written as
$
\| B \lo 1 ( B \lo 1 \trans B \lo 1) \inv B \lo 1 \trans - B \lo 2 ( B \lo 2 \trans B \lo 2)\inv B \lo 2 \trans \| \lo {\mathrm{F}}.
$
This distance is coordinate-free, as it is invariant to the basis matrices involved.

To facilitate the comparison, we also include the benchmark error, which is set as the expectation of the above distance when $B \lo 1$ is taken as any basis matrix of the true central subspace, and entries of  $B \lo 2$ are generated randomly from i.i.d. $N(0,1)$. This expectation is computed by Monte Carlo with 1000 repeats.
}

\subsection{Scenario \RN{1}: Fr\'echet SDR for distributions}

Let $(\Omega_Y, d_{\mathrm{W}})$ be the metric space of univariate distributions endowed with Wasserstein metric $d_{\mathrm{W}}$, as described in Section \ref{sec: important case}. The construction of the ensembles requires computing the Wasserstein distances $d_{\mathrm{W}}(Y_i,Y_j)$ for $i,j=1,\dots,n$. However, the distributions $Y_1,\dots, Y_n$ are usually not fully observed in practice, which means we need to estimate them in the implementation of the proposed methods. There are multiple ways to do so, such as by estimating the c.d.f.'s, the quantile functions \citep{parzen1979nonparametric}, or the p.d.f's \citep{petersen2016functional,chen2021wasserstein}. For computation simplicity, we use the Wasserstein distances between the empirical measures. Specifically, suppose we observe $(X_1, \{W_{1j}\}_{j=1}^{m_1}),\dots,(X_n, \{W_{nj}\}_{j=1}^{m_n})$, where $\{W_{ij}\}_{j=1}^{m_i}$ are independent samples from the distribution $Y_i$. Let $\hat{Y}_i$ be the empirical measure ${m_i}^{-1}\sum_{j=1}^{m_i}\delta_{W_{ij}}$, where $\delta_a$ is the Dirac measure at $a$, then we estimate $d_{\mathrm{W}}(Y_i,Y_k)$ by $d_{\mathrm{W}}(\hat{Y_i},\hat{Y}_k)$. For the theoretical justification, see \citet{fournier2015rate} and \citet{lei2020convergence}. For simplicity, we assume the sample sizes $m_{1},\dots, m\lo n$ to be the same and denote the common sample size by $m$. Then the distance between empirical measures $\hat{Y}\lo i$ and $\hat{Y}\lo k$ is a simple function of the order statistics:
$
    d_{\mathrm{W}}(\hat{Y}_i,\hat{Y}_k)=\Bigl\{\sum_{j=1}^m(W_{i(j)}-W_{k(j)})^2\Bigr\}^{1/2},
$
where $W_{i(j)}$ is the $j$-th order statistics of the sample $W_{i1}\dots, W_{im}$.

Let $\beta_1\trans = (1, 1, 0, \dots , 0)$, $\beta_2\trans = (0,\dots , 0, 1, 1)$, $\beta_3\trans=(1,2,0,\dots, 0,2)$ and $\beta_4\trans=(0,0,1,2,2,\dots,0)$. To generate univariate distributional response $Y$, we let $Y = N( \mu \lo Y, \sigma \lo Y \hi 2)$, where $\mu \lo Y$ and $\sigma \lo Y\hi 2$ are random variables dependent on $X$, and $\sigma \lo Y > 0$ almost surely, defined by the following models:
\begin{enumerate}
    \item[\textbf{\RN{1}-1}]: $\mu_Y|X\sim N(\exp(\beta_1\trans X), 1)$ and $\sigma_Y=1$.
    \item[\textbf{\RN{1}-2}]: $\mu_Y|X\sim N(\exp(\beta_1\trans X), 1)$ and $\sigma\lo Y = 10\inv\cdot 1\{\varsigma(X)< 10\inv\} + \varsigma(X)\cdot 1\{10\inv\le\varsigma(X)\le 10\} +  10\cdot 1\{\varsigma(X)> 10\}$ where $\varsigma(X)=\exp(\beta_2\trans X)$.
    {
    \item[\textbf{\RN{1}-3}]: $\mu_Y|X\sim N(3(\beta_3\trans X), 0.5^2)$ and $\sigma_Y|X= \text{Gamma} ((2+2\beta_4\trans X)^2/\nu, \nu/(2+2\beta_4\trans X))$ with truncated range $(10^{-1},10)$ and $\nu=0.5$.
    \item[\textbf{\RN{1}-4}]: $\mu_Y|X\sim N(3\sin(\beta_3\trans X), 0.5^2)$ and $\sigma_Y|X= \text{Gamma} ((2+2\beta_4\trans X)^2/\nu, \nu/(2+2\beta_4\trans X))$ with truncated range $(10^{-1},10)$ and $\nu=0.5$.}
\end{enumerate}
{ To generate the predictor $X$, we consider both the scenarios where Assumption \ref{assumption: lcm and ccv} is satisfied and violated. Specifically, for Model \RN{1}-1 and \RN{1}-2, $X$ is generated by the following two scenarios:
\begin{enumerate}
\item[(a)] $X \sim N(0,1)$; in this case both LCM and CCV in Assumption \ref{assumption: lcm and ccv} are satisfied;
\item[(b)] we first generate $U \lo 1, \ldots, U \lo p$ from the $\mathbb{AR}(1)$ model with mean 0 and covariance matrix $\Sigma = ( 0.5 \hi {|i-j|}) \lo {i,j}$, and then generate $X$ by $(\sin (U \lo 1), | U \lo 2|, U \lo 3, \ldots, U \lo p)$. For this model, both  LCM and CCS are violated.
\end{enumerate}
For Model \RN{1}-3 and \RN{1}-4, we generate $X = \Phi (U)$, where $\Phi$ is the multivariate c.d.f. of $N(0, I \lo p)$ and $U$ is generated as in (b). For this model, both  LCM and CCS are violated.
}

\cite{ying2020fr} considered similar models to Model \RN{1}-1 and Model \RN{1}-2. In Model \RN{1}-3 and Model \RN{1}-4, the error depends on $X$, which means the Fr\'echet central subspace contains direction out of the conditional Fr\'echet mean function. For Model \RN{1}-1, $B_0=\beta_1$ and $d_0=1$; for Model \RN{1}-2, $B_0=(\beta_1,\beta_2)$ and $d_0=2$; and for Models \RN{1}-3, \RN{1}-4, $B_0=(\beta_3,\beta_4)$ and $d_0=2$. In the simulation, we first generate $X \lo 1, \ldots, X \lo n$, then  generate $(\mu \lo {Y\lo 1}, \sigma \lo {Y \lo 1}), \ldots, (\mu \lo {Y\lo n}, \sigma \lo {Y \lo n})$. For each $i=1, \ldots, n$, we then generate $W \lo {i1}, \ldots, W \lo {im}$ independently from $N(\mu \lo {Y \lo i}, \sigma \lo {Y \lo i} \hi 2)$. We take $(n,p) = (200,10), (400, 20)$ and $m = 100$.

We compare performances of the CMS ensemble methods and CS ensemble methods described in Section \ref{sec:fsdr}, including FOLS, FPHD, FIHT, FSIR, FSAVE, FDR, and FOPG (with refinement). { We first implement the predictor augmentation (PA) method to estimate the dimension of the Fr\'echet central subspace. Then with estimated $\hat d$, we evaluate the estimation error.}  For FOPG, the number of iterations is set as 5, which is large enough to guarantee numerical convergence. For FSIR, FSAVE, the number of slices is chosen as $\lfloor n/2p\rfloor$; for FDR, the number of slices is chosen as $\lfloor n/6p\rfloor$. We also implement the weighted inverse regression ensemble (WIRE) method proposed by \cite{ying2020fr} for comparison. We repeat the experiments 100 times and summarize the proportion of correct identification of order and the mean and standard deviation of estimation error in Table \ref{table:dist_result} and Table \ref{table:dist_result2}. A smaller distance indicates a more accurate estimate, and the estimate with the smallest distance is shown in boldface. The benchmark distances are shown at the bottom of the table.

\begin{table}[ht!]
\begin{center}
\resizebox{0.9\columnwidth}{!}{
\begin{tabular}{*{10}{c}}
\hline
Model & $(p,n)$  & FOLS & FPHD & FIHT & FSIR & FSAVE & FDR & FOPG & WIRE\\
\hline
\multirow{6}{*}{\RN{1}-1-(a)}
    &   & 100\% & 87\% & 100\% & 98\% & 75\% & 87\% & 100\% & 100\% \\
    & (10,200) & 0.343 & 0.667 & 0.349 & 0.271 & 0.561 & 0.407 & \textbf{0.167} & 0.235 \\
    &   & (0.088) & (0.229) & (0.089) & (0.128) & (0.316) & (0.261) & (0.057) & (0.054) \\
    \cline{3-10}
    &   & 100\% & 84\% & 100\% & 100\% & 83\% & 90\% & 100\% & 100\% \\
    & (20,400) & 0.359 & 0.705 & 0.365 & 0.262 & 0.511 & 0.393 & \textbf{0.220} & 0.249 \\
    &   & (0.073) & (0.214) & (0.073) & (0.044) & (0.27) & (0.225) & (0.052) & (0.041) \\
    \cline{2-10}
\multirow{6}{*}{\RN{1}-1-(b)}
    &   & 94\% & 95\% & 93\% & 86\% & 75\% & 91\% & 100\% & 86\% \\
    & (10,200) & 0.415 & 0.663 & 0.437 & 0.33 & 0.504 & 0.338 & \textbf{0.135} & 0.3 \\
    &   & (0.18) & (0.155) & (0.187) & (0.283) & (0.341) & (0.228) & (0.036) & (0.294) \\
    \cline{3-10}
    &   & 95\% & 95\% & 96\% & 89\% & 73\% & 95\% & 99\% & 84\% \\
    & (20,400) & 0.402 & 0.667 & 0.415 & 0.303 & 0.511 & 0.297 & \textbf{0.196} & 0.314 \\
    &   & (0.156) & (0.141) & (0.142) & (0.254) & (0.329) & (0.176) & (0.089) & (0.309) \\
    \cline{2-10}
\multirow{6}{*}{\RN{1}-2-(a)}
    &   & 100\% & 57\% & 100\% & 99\% & 97\% & 100\% & 100\% & 100\% \\
    & (10,200) & 0.413 & 1.048 & 0.416 & 0.387 & 0.537 & 0.418 & \textbf{0.253} & 0.316 \\
    &   & (0.092) & (0.206) & (0.092) & (0.101) & (0.161) & (0.099) & (0.054) & (0.056) \\
    \cline{3-10}
    &   & 100\% & 48\% & 100\% & 100\% & 99\% & 100\% & 100\% & 100\% \\
    & (20,400) & 0.419 & 1.131 & 0.423 & 0.367 & 0.552 & 0.433 & \textbf{0.290} & 0.318 \\
    &   & (0.072) & (0.175) & (0.073) & (0.05) & (0.1) & (0.062) & (0.051) & (0.039) \\
    \cline{2-10}
\multirow{6}{*}{\RN{1}-2-(b)}
    &   & 100\% & 56\% & 100\% & 97\% & 98\% & 100\% & 99\% & 99\% \\
    & (10,200) & 0.552 & 1.135 & 0.558 & 0.477 & 0.627 & 0.504 & \textbf{0.288} & 0.379 \\
    &   & (0.111) & (0.143) & (0.109) & (0.147) & (0.139) & (0.107) & (0.097) & (0.11) \\
    \cline{3-10}
    &   & 100\% & 56\% & 100\% & 100\% & 100\% & 100\% & 100\% & 100\% \\
    & (20,400) & 0.570 & 1.161 & 0.576 & 0.469 & 0.658 & 0.534 & \textbf{0.337} & 0.391 \\
    &   & (0.072) & (0.121) & (0.072) & (0.066) & (0.089) & (0.07) & (0.049) & (0.054) \\
   \hline
\end{tabular}
}
\end{center}
\caption{The percentages of correct order determination, and the mean (standard deviation) of estimation error as measured by $\|P_{B_0}-P_{\hat{B}}\|_{\mathrm{F}}$ for Models \RN{1}-1 and \RN{1}-2 with settings (a) and (b); the benchmark for Model \RN{1}-1 with $p = 10, 20$ are $1.334, 1.373$ respectively, and for Model \RN{1}-2 with $p = 10, 20$ are $1.785, 1.893$, respectively. The bold-faced number indicates the best performer.}
\label{table:dist_result}
\end{table}%

\begin{table}[ht!]
\begin{center}
\resizebox{0.9\columnwidth}{!}{
\begin{tabular}{*{10}{c}}
\hline
Model & $(p,n)$  & FOLS & FPHD & FIHT & FSIR & FSAVE & FDR & FOPG & WIRE\\
\hline
\multirow{6}{*}{\RN{1}-3}
    &   & 100\% & 13\% & 100\% & 100\% & 75\% & 99\% & 100\% & 93\% \\
    & (10,200) & 0.269 & 1.150 & 0.269 & 0.342 & 0.741 & 0.408 & \textbf{0.242} & 0.311 \\
    &   & (0.05) & (0.11) & (0.051) & (0.067) & (0.303) & (0.117) & (0.047) & (0.199) \\
  \cline{3-10}
    &   & 100\% & 11\% & 100\% & 100\% & 96\% & 100\% & 100\% & 100\% \\
    & (20,400) & 0.279 & 1.165 & 0.279 & 0.333 & 0.674 & 0.407 & \textbf{0.259} & 0.273 \\
    &   & (0.046) & (0.092) & (0.046) & (0.061) & (0.177) & (0.071) & (0.042) & (0.045) \\
  \cline{2-10}
\multirow{6}{*}{\RN{1}-4}
    &   & 100\% & 39\% & 100\% & 100\% & 69\% & 99\% & 99\% & 67\% \\
    & (10,200) & 0.383 & 1.303 & 0.384 & 0.431 & 0.897 & 0.535 & \textbf{0.311} & 0.583 \\
    &   & (0.08) & (0.172) & (0.08) & (0.104) & (0.248) & (0.136) & (0.104) & (0.322) \\
  \cline{3-10}
    &   & 100\% & 47\% & 100\% & 100\% & 86\% & 99\% & 100\% & 95\% \\
    & (20,400) & 0.408 & 1.349 & 0.409 & 0.443 & 0.843 & 0.536 & \textbf{0.328} & 0.419 \\
    &   & (0.072) & (0.152) & (0.072) & (0.078) & (0.234) & (0.106) & (0.048) & (0.156) \\
   \hline
\end{tabular}
}
\end{center}
\caption{The percentages of correct order determination, and the mean (standard deviation) of estimation error as measured by $\|P_{B_0}-P_{\hat{B}}\|_{\mathrm{F}}$ for Models \RN{1}-3 and \RN{1}-4; the benchmark for Model \RN{1}-3 and \RN{1}-4 with $p = 10, 20$ are $1.787, 1.895$ respectively. The bold-faced number indicates the best performer.}
\label{table:dist_result2}
\end{table}%

{
For Model \RN{1}-1 and Model \RN{1}-2,  the best performer FOPG achieves 100\% correct order determination percentage and enjoys the smallest estimation error. The moment-based ensemble methods are slightly less accurate than FOPG. Compared with the benchmark, all methods  can successfully estimate the true central subspace except FPHD. Compared to the results from predictor settings (a) and (b), we see that most moment-based methods and inverse-regression-based methods have larger estimation error and less percentage of correct order determination under setting (b), but FOPG, which is free from the elliptical assumption of predictors, still give the most precise estimation. Overall, the correlation between predictors and non-ellipticity does not affect the results much compared with the benchmark error. In Model \RN{1}-3 and Model \RN{1}-4, there exist directions in the central subspace but not contained in the conditional Fr\'echet mean. In these cases, the best performer is still FOPG.}

\begin{figure}[ht!]
     \centering
     \begin{subfigure}[b]{0.3\textwidth}
         \centering
         \includegraphics[width=\textwidth]{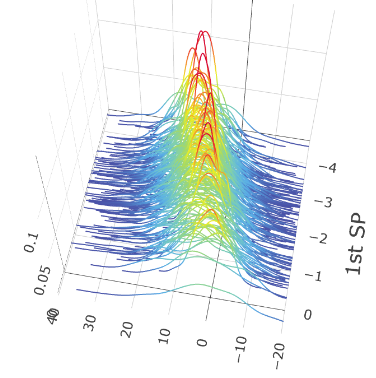}
         \caption{}
     \end{subfigure}
    \begin{subfigure}[b]{0.3\textwidth}
          \centering
          \includegraphics[width=\textwidth]{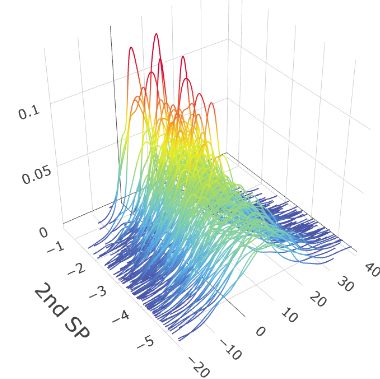}
          \caption{}
      \end{subfigure}
      \begin{subfigure}[b]{0.3\textwidth}
          \centering
          \includegraphics[width=\textwidth]{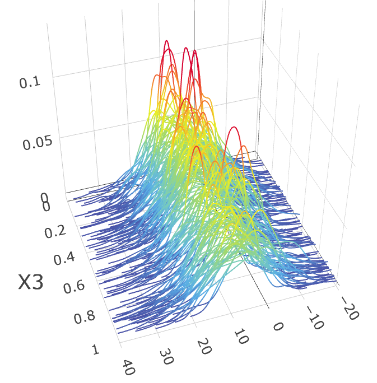}
          \caption{}
      \end{subfigure}
      \begin{subfigure}[b]{0.05\textwidth}
          \centering
          \includegraphics[width=\textwidth]{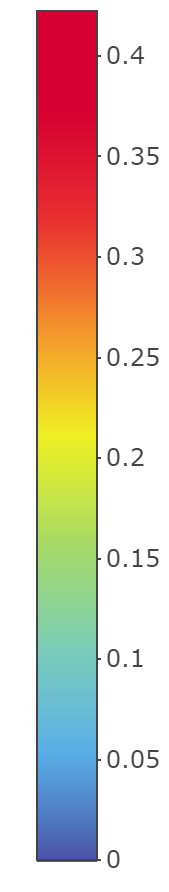}
          \caption*{}
      \end{subfigure}
      \caption{3D plots of the distributional responses versus (a) first sufficient predictor $\hat\beta \lo 1\trans X$; (b) second sufficient predictor $\hat\beta \lo 2\trans X$ and (c)  $X_{3}$ for Model \RN{1}-2 with $(n, p) = (200, 10)$}
        \label{fig:sp_dist}
\end{figure}

We also show the plots of $Y$ versus the sufficient predictors obtained by FOPG for Model \RN{1}-3. From Figure \ref{fig:sp_dist}, we see a strong relation between $Y$ and the first two estimated sufficient predictors $\hat\beta\lo 1\trans X$ and $\hat\beta\lo 2\trans X$, compared with $Y$ versus individual predictor $X_{3}$. We also observe that the first sufficient predictor captures the location variation and the second sufficient predictor captures the scale variation.
\vspace{-0.2in}
\subsection{Scenario \RN{2}: Fr\'echet SDR for positive definite matrices}
Let $\Omega_Y$ be the space $\Sym^+(r)$ endowed with Frobenius distance $d_{\mathrm{F}}(Y_1,Y_2)=\|Y_1-Y_2\|_{\mathrm{F}}$. To accommodate the anatomical intersubject variability, \cite{schwartzman2006random} introduced the symmetric matrix variate Normal distributions. We adopt this distribution to construct the regression model with correlation matrix response. We say that $Z\in\mathrm{Sym}(r)$ has the \emph{standard symmetric matrix variate Normal distribution} $N_{rr}(0;I_r)$ if it has density
$
    \varphi(Z)=(2\pi)^{-q/2}\exp(-\tr(Z)^2/2)
$
with respect to Lebesgue measure on $\Rbb^{p(p+1)}$. As pointed out in \cite{schwartzman2006random}, this definition is equivalent to a symmetric matrix with independent $N(0,1)$ diagonal elements and $N(0,1/2)$ off-diagonal elements. We say $Y\in\Sym(r)$ has symmetric matrix variate Normal distribution $N_{rr}(M;\Sigma)$ if $Y=GZG^T +M$ where $M\in\Sym(r)$, $G\in\Rbb(r\times r)$ is a non-singular matrix, and $\Sigma=G^TG$. As a special case, we say $Y\in\Sym(r)\sim N_{rr}(M;\sigma^2)$ if $Y=\sigma Z +M$.

{ We generate predictors $X$ as in settings (a) and (b) in of Scenario \RN{2}.} We generate $\log(Y)$ following $N_{dd}(\log\{D(X)\},0.25)$, where $\log(\cdot)$ is the matrix logarithm defined in Section \ref{sec: important case}, and $D(X)$ is specified by the following models:
\begin{enumerate}
    \item[] \textbf{\RN{2}-1}:$
    D(X)=\begin{pmatrix}
    1& \rho(X) \\
    \rho(X)& 1\\
    \end{pmatrix}
    $,
    where $\rho(X)=[\exp(\beta_1\trans X)-1]/[\exp(\beta_1\trans X)+1]$.
    \item[] \textbf{\RN{2}-2}: $
    D(X)=\begin{pmatrix}
    1& \rho_1(X) & \rho_2(X)\\
    \rho_1(X)& 1& \rho_1(X)\\
    \rho_2(X)& \rho_1(X) & 1
    \end{pmatrix}
   $,
    where $\rho_1(X)=0.4[\exp(\beta_1\trans X)-1]/[\exp(\beta_1\trans X)+1]$ and $\rho_2(X)=0.4\sin(\beta_3\trans X)$.
\end{enumerate}
In Model \RN{2}-1, $B_0=\beta_1$ and $d_0=1$; in Model \RN{2}-2, $B_0=(\beta_1,\beta_2)$ and $d_0=2$. We note that $D(x)$ is not necessarily the Fr\'echet conditional mean of $Y$ given $X$, but it still measures the central tendency of the conditional distribution $Y|X$. {  We also compare performances of the CMS ensemble methods and CS ensemble methods, with $(n,p) = (200,10),(400,20)$. The experiments are repeated 100 times. The proportion of correct identification of order and the means and standard deviations of estimation errors are summarized in Table \ref{table:spd}.}
\begin{table}[ht!]
\begin{center}
\resizebox{0.9\columnwidth}{!}{
\begin{tabular}{*{10}{c}}
\hline
Model & $(p,n)$  & FOLS & FPHD & FIHT & FSIR & FSAVE & FDR & FOPG & WIRE\\
\hline
\multirow{6}{*}{\RN{2}-1-(a)}
    &   & 100\% & 68\% & 100\% & 100\% & 91\% & 99\% & 100\% & 100\% \\
    & (10,200) & 0.153 & 0.836 & 0.153 & 0.158 & 0.264 & 0.159 & \textbf{0.150} & \textbf{0.150} \\
    &   & (0.043) & (0.296) & (0.043) & (0.037) & (0.245) & (0.095) & (0.04) & (0.041) \\
    \cline{3-10}
    &   & 100\% & 69\% & 100\% & 100\% & 89\% & 99\% & 100\% & 100\% \\
    & (20,400) & 0.160 & 0.905 & 0.160 & 0.169 & 0.283 & 0.165 & \textbf{0.151} & 0.158 \\
    &   & (0.029) & (0.267) & (0.029) & (0.026) & (0.262) & (0.088) & (0.026) & (0.028) \\
    \cline{2-10}
\multirow{6}{*}{\RN{2}-1-(b)}
    &   & 97\% & 64\% & 97\% & 95\% & 53\% & 64\% & 93\% & 98\% \\
    & (10,200) & 0.248 & 1.017 & 0.249 & 0.276 & 0.726 & 0.525 & 0.251 & \textbf{0.218} \\
    &   & (0.154) & (0.276) & (0.155) & (0.186) & (0.342) & (0.387) & (0.219) & (0.133) \\
    \cline{3-10}
    &  & 100\% & 54\% & 100\% & 97\% & 43\% & 63\% & 98\% & 100\% \\
    & (20,400) & 0.233 & 1.117 & 0.233 & 0.252 & 0.779 & 0.533 & 0.216 & \textbf{0.213} \\
    &   & (0.049) & (0.247) & (0.049) & (0.144) & (0.349) & (0.388) & (0.121) & (0.045) \\
    \cline{2-10}
\multirow{6}{*}{\RN{2}-2-(a)}
    &   & 100\% & 19\% & 100\% & 99\% & 78\% & 99\% & 100\% & 100\% \\
    & (10,200) & 0.295 & 1.185 & 0.295 & 0.310 & 0.619 & 0.371 & \textbf{0.149} & 0.293 \\
    &   & (0.085) & (0.139) & (0.085) & (0.113) & (0.293) & (0.133) & (0.034) & (0.084) \\
    \cline{3-10}
    &   & 100\% & 24\% & 100\% & 100\% & 81\% & 100\% & 100\% & 100\% \\
    & (20,400) & 0.312 & 1.231 & 0.312 & 0.325 & 0.626 & 0.382 & \textbf{0.181} & 0.311 \\
    &   & (0.065) & (0.155) & (0.065) & (0.064) & (0.233) & (0.084) & (0.032) & (0.065) \\
   \cline{2-10}
\multirow{6}{*}{\RN{2}-2-(b)}
    &   & 100\% & 32\% & 100\% & 100\% & 60\% & 76\% & 100\% & 100\% \\
    & (10,200) & 0.675 & 1.461 & 0.675 & 0.697 & 1.193 & 0.920 & \textbf{0.245} & 0.671 \\
    &   & (0.183) & (0.195) & (0.183) & (0.186) & (0.246) & (0.249) & (0.071) & (0.182) \\
    \cline{3-10}
    &   & 100\% & 43\% & 100\% & 100\% & 57\% & 90\% & 100\% & 100\% \\
    & (20,400) & 0.697 & 1.492 & 0.697 & 0.714 & 1.253 & 0.936 & \textbf{0.331} & 0.694 \\
    &   & (0.136) & (0.182) & (0.136) & (0.13) & (0.206) & (0.208) & (0.075) & (0.137) \\
   \hline
\end{tabular}
}
\end{center}
\caption{Mean($\pm$ standard deviation) of estimation error measured by $\|P_{B_0}-P_{\hat{B}}\|_{\mathrm{F}}$ for different methods for Scenario \RN{2}. The benchmark for Model \RN{2}-1 with $p = 10, 20$ are $1.334, 1.373$ respectively, for Model \RN{2}-2 with $p = 10, 20$ are $1.785, 1.893$, respectively. The bold-faced number indicates the best performer. }
\label{table:spd}
\end{table}%

\begin{figure}[ht!]
     \centering
     \begin{subfigure}[b]{0.35\textwidth}
         \centering
         \includegraphics[width=\textwidth]{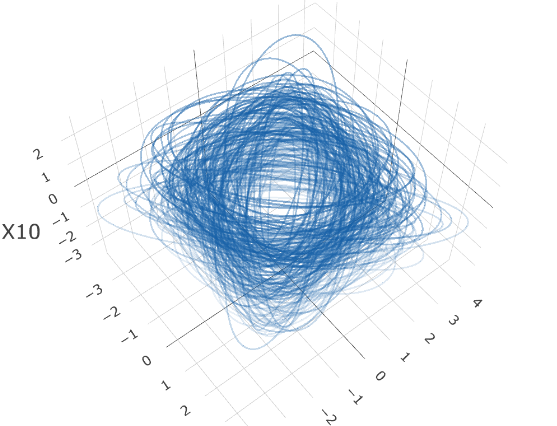}
         \caption{}
     \end{subfigure}
     \hspace{0.2in}
     \begin{subfigure}[b]{0.35\textwidth}
         \centering
         \includegraphics[width=\textwidth]{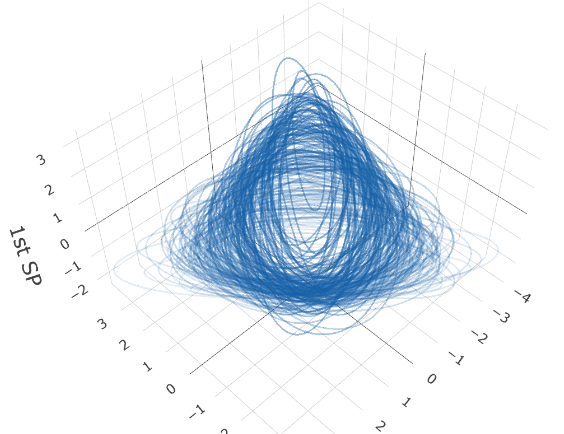}
         \caption{}
     \end{subfigure}
      \caption{Ellipsoidal plots of the SPD matrix response versus the FOPG predictor $\hat\beta_1\trans X$ and  $X_{10}$ using Model \RN{2}-1-(a) with $(n=200, p=10)$. Each horizontal ellipse is an Ellipsoidal representation of  an SPD matrix, and the vertical axis is the value of (a) $\hat\beta_1\trans X$; (b) $X_{10}$.}
        \label{fig:sp_spd}
\end{figure}

We conclude that all ensemble methods give reasonable estimation except FPHD.  FOPG performs best in all settings except Model \RN{2}-1-(b). To illustrate the relation between the response and estimated sufficient predictor $\hat\beta_1\trans X$, we adopt the ellipsoidal representation of SPD matrices. Each $A\in\Sym^+(d)$ can be associated with an ellipsoid centered at the origin $\mathcal{E}_A=\{x:x\trans A^{-1} x \le 1\}$. Figure \ref{fig:sp_spd} plots the responses ellipsoid versus the estimated sufficient predictor in panel (a), compared with the responses versus predictor $X_{10}$ for Model \RN{2}-1-(a). We can tell a clear pattern of change in shape and rotation of response ellipsoids versus $\hat\beta_1\trans X$.

\section{Application to the Human Mortality Data}\label{sec:realdata}
This section presents an application concerning  human life  spans. Another application concerning intracerebral hemorrhage is presented in Section S.6 of the Supplementary Material.

Compared with summary statistics such as the crude death rate, viewing the entire age-at-death distributions as data objects gives us a more comprehensive understanding of human longevity and health conditions. Mortality distributions are affected by many factors, such as economics, the health care system, as well as social and environmental factors. To investigate the potential factors that are related to the mortality distributions across different countries, we collect nine predictors listed below, covering demography, economics, labor market, nutrition, health, and environmental factors in 2015:
(1) Population Density: population per square Kilometer; (2) Sex Ratio: number of males per 100 females in the population;
    (3) Mean Childbearing Age: the average age of mothers at the birth of their children;
    (4) Gross Domestic Product (GDP) per Capita;
    (5) Gross Value Added (GVA) by Agriculture: the percentage of agriculture, hunting, forestry, and fishing activities of gross value added,
    (6) Consumer price index: treat 2010 as the base year;
    (7) Unemployment Rate;
    (8) Expenditure on Health (percentage of GDP);
    (9) Arable Land (percentage of total land area).
The data are collected from United Nation Databases (http://data.un.org/) and UN World Population Prospects 2019 Databases (https://population.un.org/wpp/Download). For each country and age, the life table contains the number of deaths $d(x,n)$ aggregated every 5 years. We treat these data as histograms of the number of deaths at age, with bin widths equal to 5 years. We smooth the histograms for the 162 countries in 2015 using the `frechet' package available at (https://cran.r-project.org/web/packages/frechet/index.html) to obtain smoothed probability density functions. We then calculate the Wasserstein distances between them.

\begin{figure}[ht]
     \centering
     \begin{subfigure}[b]{0.35\textwidth}
         \centering
         \includegraphics[width=\textwidth]{Graph/mortality_sp1.png}
         \caption{}
     \end{subfigure}
     \begin{subfigure}[b]{0.55\textwidth}
         \centering
         \includegraphics[width=\textwidth]{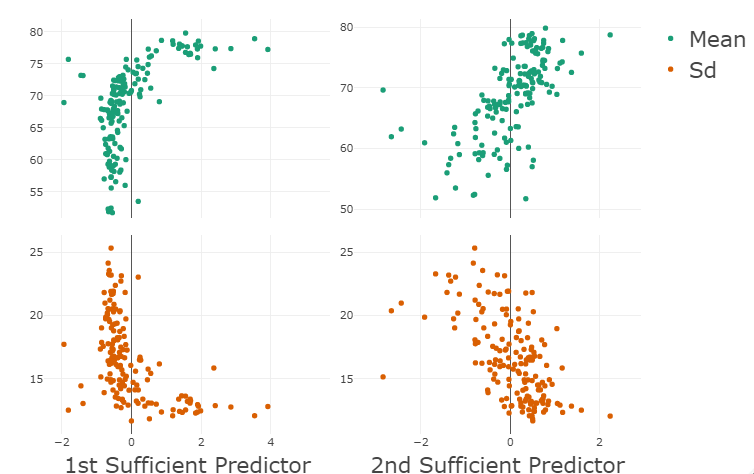}
         \caption{}
     \end{subfigure}

      \caption{(a) Mortality distributions versus the first sufficient predictors; (b) Mean and standard deviation of mortality distributions versus the first two sufficient predictors.}
        \label{fig:mort_dis_1sp}
\end{figure}

{  We use Gaussian kernel $\kappa(y, y') = \exp(-\gamma d\hi2\lo W(y, y'))$, where $\gamma$ is taken according to \eqref{eq: kernel_para} in Section 6.1. We standardize all covariates separately, then use the predictor augmentation method combined with FOPG to estimate the dimension of the Fr\'echet central subspace. The estimated dimension of the Fr\'echet central subspace is $3$. The first three directions obtained by FOPG are
\begin{align*}
   &\hat\beta_1 = (0.416, -0.424,  0.114,  0.770,  0.027, -0.146, -0.020, 0.130, -0.053)\trans \\
   &\hat\beta_2 = (0.186,  0.155, -0.159, -0.135, -0.576, -0.714, -0.083, 0.198, 0.096)\trans\\
   &\hat\beta_3 = (-0.108, 0.498,  0.507,  0.135,  0.487, -0.403,  0.139, 0.211,  0.045)\trans.
\end{align*}
}
A plot of mortality densities versus the first sufficient predictor $\hat \beta \lo 1 \hi T X$ is shown in Figure \ref{fig:mort_dis_1sp}(a).
Clear and useful patterns emerge from Figure \ref{fig:mort_dis_1sp}(a): the mode of the mortality distribution shifts from right to left (with left indicating a longer life span) as the value of the first sufficient predictor increases. Moreover,  there is a significant uptick at the right-most end  as the first sufficient predictor decreases, indicating high infant mortality. Meanwhile, the loadings of the first sufficient predictor are strongly positive for the  GDP per capita, which indicates the levels of economic development and health care of a country, with larger values associated with more  developed countries and smaller values associated with less developed countries. From Figure \ref{fig:mort_dis_1sp}(b), we see that the mean of the mortality distribution increases and the standard deviation decreases with the value of the first sufficient predictor. This also makes sense: the mean life span increases with the level of development, consistent with Figure \ref{fig:mort_dis_1sp}(a). The standard deviation decreases with the first predictor because, as the development level increases, the life span is increasingly concentrated on the high values. Moreover, the high mortality in the lower region of the first sufficient predictor also contributes to the larger standard deviation in this region. The plots of mean and standard deviation versus the second sufficient predictor in Figure \ref{fig:mort_dis_1sp}(b) also indicate an increase in mean and a decrease in standard deviation as the value of the second sufficient predictor increases.

\section{Discussion}\label{sec:discuss}
In the classical regression setting, sufficient dimension reduction has been used as a tool for exploratory data analysis, regression diagnostics, and a mechanism to overcome the curse of dimensionality in regression. As a regression tool, it can help us to effectively treat collinearity in the predictor, detect heteroscedasticity in the response, find the most important linear combinations of predictors, and understand the general shape of the regression surface without fitting an elaborate regression model. All the above items can be revealed by the sufficient plot, where the response is plotted against the sufficient predictors obtained by SDR.  Also, because the direction of a vector is an easier target  to estimate than the vector itself, it helps us to mitigate the effect of the curse of dimensionality usually accompanying  high-dimensional regression. In the same vein, it is easier to estimate a subspace than a specific set of vectors that span the subspace.

Although regression with a metric-space valued random object is a new problem, as a regression problem, it shares the same set of  issues, such as the need for exploratory analysis before regression, for model diagnostics after the regression, and for mitigating the curse of dimensionality. As shown in Figure 1 in the paper, the first sufficient predictor clearly reveals useful information about a general trend of  mortality distributions among countries. At the lower end of the sufficient predictor, the mortality distribution shifts towards higher longevity, whereas at the high end of the sufficient predictor, the mortality distribution shifts towards low longevity, with a visible  uptick near age 0, which is caused by infant mortality.

The proposed methodology is very flexible and versatile: it can be used to turn any existing SDR method into one that can deal with the metric-space-valued response variable. Furthermore, it applies to any separable and complete metric space of negative type with an explicit CMS ensemble. It significantly broadens the current field of sufficient dimension reduction and provides a useful set of tools for Fr\'echet regression.

{\footnotesize
\bibliographystyle{agsm}
\bibliography{main_arxiv}

@article{dubey2019frechet,
  title={Fr{\'e}chet analysis of variance for random objects},
  author={Dubey, Paromita and M{\"u}ller, Hans-Georg},
  journal={Biometrika},
  volume={106},
  number={4},
  pages={803--821},
  year={2019},
  publisher={Oxford University Press}
}

@article{petersen2019frechet,
  title={Fr{\'e}chet regression for random objects with Euclidean predictors},
  author={Petersen, Alexander and M{\"u}ller, Hans-Georg},
  journal={The Annals of Statistics},
  volume={47},
  number={2},
  pages={691--719},
  year={2019},
  publisher={Institute of Mathematical Statistics}
}

@article{ying2020fr,
  title={Fr\'echet Sufficient Dimension Reduction for Random Objects},
  author={Ying, Chao and Yu, Zhou},
  journal={Biometrika, in press},
  volume={},
  number={},
  pages={},
  year={2022}
}

@article{cook1996graphics,
  title={Graphics for regressions with a binary response},
  author={Cook, R Dennis},
  journal={J. Amer. Statist. Assoc},
  volume={91},
  pages={983--992},
  year={1996},
  publisher={Taylor \& Francis}
}

@article{cook2002dimension,
  title={Dimension reduction for conditional mean in regression},
  author={Cook, R Dennis and Li, Bing},
  journal={The Annals of Statistics},
  volume={30},
  number={2},
  pages={455--474},
  year={2002},
  publisher={Institute of Mathematical Statistics}
}

@article{yin2011sufficient,
  title={Sufficient dimension reduction based on an ensemble of minimum average variance estimators},
  author={Yin, Xiangrong and Li, Bing},
  journal={The Annals of Statistics},
  volume={39},
  number={6},
  pages={3392--3416},
  year={2011},
  publisher={Institute of Mathematical Statistics}
}

@INPROCEEDINGS{Christmann10universalkernels,
    author = {Andreas Christmann and Ingo Steinwart},
    title = {Universal kernels on non-standard input spaces},
    booktitle = {in Advances in Neural Information Processing Systems},
    year = {2010},
    pages = {406--414}
}

@article{ambrosio2004gradient,
  title={Gradient flows with metric and differentiable structures, and applications to the Wasserstein space},
  author={Ambrosio, Luigi and Gigli, Nicola and Savar{\'e}, Giuseppe},
  journal={Atti Accad. Naz. Lincei Cl. Sci. Fis. Mat. Natur. Rend. Lincei (9) Mat. Appl},
  volume={15},
  number={3-4},
  pages={327--343},
  year={2004}
}

@article{bigot2017geodesic,
  title={Geodesic PCA in the Wasserstein space by convex PCA},
  author={Bigot, J{\'e}r{\'e}mie and Gouet, Ra{\'u}l and Klein, Thierry and L{\'o}pez, Alfredo},
  journal={Annales de l'Institut Henri Poincar{\'e}, Probabilit{\'e}s et Statistiques},
  volume={53},
  pages={1--26},
  year={2017}
}

@article{xia2007constructive,
  title={A constructive approach to the estimation of dimension reduction directions},
  author={Xia, Yingcun},
  journal={The Annals of Statistics},
  volume={35},
  number={6},
  pages={2654--2690},
  year={2007},
  publisher={Institute of Mathematical Statistics}
}

@article{li2005contour,
  title={Contour regression: a general approach to dimension reduction},
  author={Li, Bing and Zha, Hongyuan and Chiaromonte, Francesca},
  journal={The Annals of Statistics},
  volume={33},
  number={4},
  pages={1580--1616},
  year={2005},
  publisher={Institute of Mathematical Statistics}
}

@article{petersen2016functional,
  title={Functional data analysis for density functions by transformation to a Hilbert space},
  author={Petersen, Alexander and M{\"u}ller, Hans-Georg},
  journal={The Annals of Statistics},
  volume={44},
  number={1},
  pages={183--218},
  year={2016},
  publisher={Institute of Mathematical Statistics}
}

@article{chen2021wasserstein,
  title={Wasserstein regression},
  author={Chen, Yaqing and Lin, Zhenhua and M{\"u}ller, Hans-Georg},
  journal={J. Amer. Statist. Assoc.},
  volume={},
  number={},
  pages={},
  year={2021},
  publisher={Taylor \& Francis}
}

@article{parzen1979nonparametric,
  title={Nonparametric statistical data modeling},
  author={Parzen, Emanuel},
  journal={J. Amer. Statist. Assoc.},
  volume={74},
  number={365},
  pages={105--121},
  year={1979},
  publisher={Taylor \& Francis}
}

@article{fournier2015rate,
  title={On the rate of convergence in Wasserstein distance of the empirical measure},
  author={Fournier, Nicolas and Guillin, Arnaud},
  journal={Probability Theory and Related Fields},
  volume={162},
  number={3-4},
  pages={707--738},
  year={2015},
  publisher={Springer}
}

@article{lei2020convergence,
  title={Convergence and concentration of empirical measures under wasserstein distance in unbounded functional spaces},
  author={Lei, Jing},
  journal={Bernoulli},
  volume={26},
  number={1},
  pages={767--798},
  year={2020},
  publisher={Bernoulli Society for Mathematical Statistics and Probability}
}

@inproceedings{feragen2015geodesic,
  title={Geodesic exponential kernels: when curvature and linearity conflict},
  author={Feragen, Aasa and Lauze, Fran{\c{c}}ois and Hauberg, Soren},
  booktitle={Proceedings of the IEEE Conference on Computer Vision and Pattern Recognition},
  pages={3032--3042},
  year={2015}
}

@inproceedings{jayasumana2013combining,
  title={Combining multiple manifold-valued descriptors for improved object recognition},
  author={Jayasumana, Sadeep and Hartley, Richard and Salzmann, Mathieu and Li, Hongdong and Harandi, Mehrtash},
  booktitle={2013 International Conference on Digital Image Computing: Techniques and Applications (DICTA)},
  pages={1--6},
  year={2013},
  organization={IEEE}
}

@inproceedings{honeine2010angular,
  title={The angular kernel in machine learning for hyperspectral data classification},
  author={Honeine, Paul and Richard, C{\'e}dric},
  booktitle={2010 2nd Workshop on Hyperspectral Image and Signal Processing: Evolution in Remote Sensing},
  pages={1--4},
  year={2010},
  organization={IEEE}
}

@book{li2018sufficient,
  title={Sufficient Dimension Reduction: Methods and Applications with R},
  author={Li, Bing},
  year={2018},
  publisher={CRC Press}
}

@article{li1989regression,
  title={Regression analysis under link violation},
  author={Li, Ker-Chau and Duan, Naihua},
  journal={Ann. Stat.},
  volume={17},
  number={3},
  pages={1009--1052},
  year={1989},
  publisher={JSTOR}
}

@article{li1991sliced,
  title={Sliced inverse regression for dimension reduction},
  author={Li, Ker-Chau},
  journal={J. Amer. Statist. Assoc},
  volume={86},
  pages={316--327},
  year={1991},
  publisher={Taylor \& Francis Group}
}

@article{li1992principal,
  title={On principal Hessian directions for data visualization and dimension reduction: Another application of Stein's lemma},
  author={Li, Ker-Chau},
  journal={Journal of the American Statistical Association},
  volume={87},
  number={420},
  pages={1025--1039},
  year={1992},
  publisher={Taylor \& Francis}
}

@article{cook1991sliced,
  title={Sliced inverse regression for dimension reduction: Comment},
  author={Cook, R Dennis and Weisberg, Sanford},
  journal={Journal of the American Statistical Association},
  volume={86},
  number={414},
  pages={328--332},
  year={1991},
  publisher={JSTOR}
}

@article{li2007directional,
  title={On directional regression for dimension reduction},
  author={Li, Bing and Wang, Shaoli},
  journal={Journal of the American Statistical Association},
  volume={102},
  number={479},
  pages={997--1008},
  year={2007},
  publisher={Taylor \& Francis}
}

@article{ye2003using,
  title={Using the bootstrap to select one of a new class of dimension reduction methods},
  author={Ye, Zhishen and Weiss, Robert E},
  journal={Journal of the American Statistical Association},
  volume={98},
  number={464},
  pages={968--979},
  year={2003},
  publisher={Taylor \& Francis}
}

@article{yu2020nonparametric,
  title={Nonparametric Estimation and Conformal Inference of the Sufficient Forecasting with a Diverging Number of Factors},
  author={Yu, Xiufan and Yao, Jiawei and Xue, Lingzhou},
  journal={J. Bus. Econ. Stat.},
  volume={40},
  number={1},
  pages={342--354},
  year={2020},
  publisher={Taylor \& Francis}
}

@article{fan2017sufficient,
  title={Sufficient forecasting using factor models},
  author={Fan, Jianqing and Xue, Lingzhou and Yao, Jiawei},
  journal={Journal of Econometrics},
  volume={201},
  number={2},
  pages={292--306},
  year={2017},
  publisher={Elsevier}
}

@article{luo2021inverse,
  title={Inverse moment methods for sufficient forecasting using high-dimensional predictors},
  author={Luo, Wei and Xue, Lingzhou and Yao, Jiawei and Yu, Xiufan},
  journal={Biometrika},
  volume={109},
  number={2},
  pages={473–-487},
  year={2021}
}

@article{xia2002adaptive,
  title={An adaptive estimation of dimension reduction space (with discussion)},
  author={Xia, Yingcun and Tong, Howell and Li, Wai Keung and Zhu, Li-Xing},
  journal={Journal of the Royal Statistical Society. Series B},
  volume={64},
  number={3},
  pages={363--410},
  year={2002},
  publisher={Wiley-Blackwell}
}

@article{ferre2003functional,
  title={Functional sliced inverse regression analysis},
  author={Ferr{\'e}, Louis and Yao, Anne-Fran{\c{c}}oise},
  journal={Statistics},
  volume={37},
  number={6},
  pages={475--488},
  year={2003},
  publisher={Taylor \& Francis}
}

@article{li2017nonlinear,
  title={Nonlinear sufficient dimension reduction for functional data},
  author={Li, Bing and Song, Jun},
  journal={The Annals of Statistics},
  volume={45},
  number={3},
  pages={1059--1095},
  year={2017},
  publisher={Institute of Mathematical Statistics}
}

@article{li2010dimension,
  title={On dimension folding of matrix-or array-valued statistical objects},
  author={Li, Bing and Kim, Min Kyung and Altman, Naomi},
  journal={The Annals of Statistics},
  volume={38},
  number={2},
  pages={1094--1121},
  year={2010},
  publisher={Institute of Mathematical Statistics}
}

@article{ferreira2013resting,
  title={Resting-state functional connectivity in normal brain aging},
  author={Ferreira, Luiz Kobuti and Busatto, Geraldo F},
  journal={Neuroscience \& Biobehavioral Reviews},
  volume={37},
  number={3},
  pages={384--400},
  year={2013},
  publisher={Elsevier}
}

@article{frechet1948elements,
  title={Les {\'e}l{\'e}ments al{\'e}atoires de nature quelconque dans un espace distanci{\'e}},
  author={Fr{\'e}chet, M.},
  journal={Annales de l'Institut Henri Poincar{\'e}},
  url={},
  volumn={10},
  pages={215–310},
  year={1948},
  publisher={Gauthier-Villars}
}

@article{dubey2020frechet,
  title={Fr{\'e}chet change-point detection},
  author={Dubey, Paromita and M{\"u}ller, Hans-Georg},
  journal={Ann. Stat.},
  volume={48},
  number={6},
  pages={3312--3335},
  year={2020},
  publisher={Institute of Mathematical Statistics}
}

@phdthesis{schwartzman2006random,
  title={Random ellipsoids and false discovery rates: Statistics for diffusion tensor imaging data},
  author={Schwartzman, Armin},
  year={2006},
  school={Stanford University}
}

@article{hsing2009rkhs,
  title={An RKHS formulation of the inverse regression dimension-reduction problem},
  author={Hsing, Tailen and Ren, Haobo},
  journal={The Annals of Statistics},
  volume={37},
  number={2},
  pages={726--755},
  year={2009},
  publisher={Institute of Mathematical Statistics}
}

@article{ding2015tensor,
  title={Tensor sliced inverse regression},
  author={Ding, Shanshan and Cook, R Dennis},
  journal={J. Multivar. Anal.},
  volume={133},
  pages={216--231},
  year={2015},
  publisher={Elsevier}
}

@article{yin2008successive,
  title={Successive direction extraction for estimating the central subspace in a multiple-index regression},
  author={Yin, Xiangrong and Li, Bing and Cook, R Dennis},
  journal={Journal of Multivariate Analysis},
  volume={99},
  number={8},
  pages={1733--1757},
  year={2008},
  publisher={Elsevier}
}

@article{eaton1986characterization,
  title={A characterization of spherical distributions},
  author={Eaton, Morris L},
  journal={J. Multivar. Anal.},
  volume={20},
  number={2},
  pages={272--276},
  year={1986},
  publisher={Elsevier}
}

@article{luo2016combining,
  title={Combining eigenvalues and variation of eigenvectors for order determination},
  author={Luo, Wei and Li, Bing},
  journal={Biometrika},
  volume={103},
  number={4},
  pages={875--887},
  year={2016},
  publisher={Oxford University Press}
}

@article{zhu2006sliced,
  title={On sliced inverse regression with high-dimensional covariates},
  author={Zhu, Lixing and Miao, Baiqi and Peng, Heng},
  journal={Journal of the American Statistical Association},
  volume={101},
  number={474},
  pages={630--643},
  year={2006},
  publisher={Taylor \& Francis}
}

@article{lee2013general,
  title={A general theory for nonlinear sufficient dimension reduction: Formulation and estimation},
  author={Lee, Kuang-Yao and Li, Bing and Chiaromonte, Francesca},
  journal={The Annals of Statistics},
  volume={41},
  number={1},
  pages={221--249},
  year={2013},
  publisher={Institute of Mathematical Statistics}
}

@article{hall1993almost,
  title={On almost linearity of low dimensional projections from high dimensional data},
  author={Hall, Peter and Li, Ker-Chau},
  journal={The Annals of Statistics},
  pages={867--889},
  year={1993},
  publisher={JSTOR}
}

@article{micchelli2006universal,
  title={Universal Kernels.},
  author={Micchelli, Charles A and Xu, Yuesheng and Zhang, Haizhang},
  journal={J. Mach. Learn. Res.},
  volume={7},
  number={12},
  pages={2651--2667},
  year={2006}
}

@article{granirer_1970, title={Probability Measures on Metric Spaces. by K. R. Parthasarathy. Academic Press, New York and London (1967). x   276 pp.}, volume={13}, DOI={10.1017/S0008439500031787}, number={2}, journal={Canadian Mathematical Bulletin}, publisher={Cambridge University Press}, author={Granirer, E. E.}, year={1970}, pages={290–291}}

@book{panaretos2020invitation,
  title={An Invitation to Statistics in Wasserstein Space},
  author={Panaretos, Victor M and Zemel, Yoav},
  year={2020},
  publisher={Springer Nature}
}

@article{li2007surrogate,
  title={On surrogate dimension reduction for measurement error regression: an invariance law},
  author={Li, Bing and Yin, Xiangrong},
  journal={The Annals of Statistics},
  volume={35},
  number={5},
  pages={2143--2172},
  year={2007},
  publisher={Institute of Mathematical Statistics}
}

@article{luo2021order,
  title={On order determination by predictor augmentation},
  author={Luo, Wei and Li, Bing},
  journal={Biometrika},
  volume={108},
  number={3},
  pages={557--574},
  year={2021},
  publisher={Oxford University Press}
}

@article{steinwart2001influence,
  title={On the influence of the kernel on the consistency of support vector machines},
  author={Steinwart, Ingo},
  journal={Journal of Machine Learning Research},
  volume={2},
  number={Nov},
  pages={67--93},
  year={2001}
}

@article{sriperumbudur2011universality,
  title={Universality, Characteristic Kernels and RKHS Embedding of Measures.},
  author={Sriperumbudur, Bharath K and Fukumizu, Kenji and Lanckriet, Gert RG},
  journal={Journal of Machine Learning Research},
  volume={12},
  number={7},
  pages={2389--2410},
  year={2011}
}
}
\end{document}